%% file: v9.tex
\documentclass[useAMS,usenatbib,galley]{mn2e}
\input mybeg.tex

\input units.tex

\usepackage{graphicx}
\usepackage{txfonts}
\usepackage{natbib}
\usepackage{wrapfig}                  
\usepackage{amssymb}
\usepackage{longtable}
\usepackage{times}
\usepackage[usenames]{color}
\usepackage{psfig}
\usepackage{soul,color}
\usepackage{subfigure}
\newcommand*{\mysub}[2]{\ensuremath{#1_{\mathrm{#2}}}}
\newcommand*{\Omegam}{\mysub{\Omega}{m}}

\newcommand*{\Omegal}{\ensuremath{\Omega_{\Lambda}}}

\newcommand*{\LCDM}{\ensuremath{\Lambda}CDM}

\newcommand*{\msolar}{\mysub{M}{\odot}}

\newcommand*{\thresh}{\mysub{P}{b} \ensuremath{< 10^{-3}}}
\newcommand*{\unthresh}{\mysub{P}{b} \ensuremath{\geq 10^{-3}}}
\newcommand*\prob{\mysub{P}{b}}
\newcommand*\Bsrc{\mysub{B}{src}}
\newcommand*\Bext{\mysub{B}{ext}}

\title[X-ray Bright AGN in Massive Clusters]
  { X-ray Bright Active Galactic Nuclei in Massive Galaxy Clusters I: Number Counts and Spatial Distribution}
\author[S. Ehlert et al.]
{S.~Ehlert,$^1$\thanks{Email:sehlert@stanford.edu.}
S.W.~Allen,$^1$
W.N.~Brandt,$^{2,3}$
Y.Q.~Xue,$^{2,3,4}$
B.~Luo,$^{2,3}$
\newauthor
A.~von der Linden,$^1$ 
A.~Mantz,$^{5,6}$
and R.G.~Morris,$^1$\\
$^1$Kavli Institute for Particle Astrophysics and Cosmology
382 Via Pueblo Mall, Stanford CA 94305, USA\\
and SLAC National Accelerator Laboratory, 2575 Sand Hill Road, Menlo Park, CA 94025, USA\\
$^2$ Department of Astronomy and Astrophysics, Pennsylvania State University, University Park, PA 16802, USA\\
$^3$ Institute for Gravitation and the Cosmos, Department of Phyiscs, Pennsylvania State University, University Park, PA 16802, USA\\
$^4$ Key Laboratory for Research in Galaxies and Cosmology, Department of Astronomy, \\
University of Science and Technology of China, Chinese Academy of Sciences, Hefei, Anhui 230026, China\\
$^5$ Kavli Institute for Cosmological Physics, 5640 S. Ellis Avenue, Chicago, IL 60637, USA\\
$^6$ Department of Astronomy and Astrophysics, University of Chicago, 5640 S. Ellis Avenue, Chicago, IL 60637, USA }
\def\cha{{\it Chandra}}
\def\ae{{\small ACIS-EXTRACT}}

\def\marx{{\small MARX}}
\def\sub{{\it Subaru}}

\def\rfive{$r_{500}$}
\def\rosat{{\it ROSAT}}
\def\rtwo{ $r_{200}$}
\def\xmm{{\it XMM-Newton}}

\def\arcsec {\hbox{$^{\prime\prime}$}} 
\def\arcmin {\hbox{$^{\prime}$}} 
\def\MACS {MACS J1931.8-2634}

\def\wav{{\small WAVDETECT}}

\def\cdfs{{\it CDFS}}
\def\cdfn{{\it CDFN}}

\begin{document}

\maketitle

\begin{abstract}
We present an analysis of the X-ray bright point source population in 43 massive clusters of galaxies observed with the \cha \ X-ray Observatory. We have constructed a catalog of 4210 rigorously selected X-ray point sources in these fields, which span a survey area of 4.2 $\deg^{2}$. This catalog reveals a clear excess of sources when compared to deep blank-field surveys, which amounts to roughly $1 \pm 0.4$ additional source per cluster, likely Active Galactic Nuclei (AGN) associated with the clusters. The excess sources are located within the overdensity radius \rfive, with the largest excess observed near the cluster centers. The average radial profile of the excess X-ray sources of the cluster are well described by a power law ($N(r) \sim r^{\beta}$) with an index of $\beta \sim -0.6$. An initial analysis using literature results on the mean profile of member galaxies in massive X-ray selected clusters indicates that the fraction of galaxies hosting X-ray AGN rises with increasing clustercentric radius, being approximately $3$ times higher near \rfive \ than in the central regions. This trend is qualitatively similar to that observed for star formation in cluster member galaxies.  
\end{abstract}

\section{Introduction}

A critical element of galaxy evolution models is the influence of the local environment.  Environmental effects are expected to be particularly pronounced in galaxy clusters, where hundreds or thousands of galaxies are in close proximity to one another, and are embedded in a hot, diffuse intracluster medium (ICM). Previous studies have shown that the ICM has clear effects on member galaxy properties: the star formation rate of galaxies in clusters is lower when compared to field galaxies, resulting in cluster members being observably redder and more elliptical \citep[e.g.][]{Dressler1980}. This suggests that the ICM is efficient at stripping cold gas from galaxies and suppressing star formation, at least since $z \sim 1$ \citep[e.g.][]{Elbaz2007}.

Theoretical and observational studies have shown that the evolution of Active Galactic Nuclei (AGN) and their host galaxies are linked to one another \citep[e.g.][]{Silk1998}. The masses of the supermassive black holes at the centers of nearby galaxies correlate tightly with the masses of their galaxy bulges \citep[e.g.][]{Magorrian1998,Gebhardt2000,Marconi2003}, suggesting a connection between their growth histories. Feedback between AGN and their host galaxies has also been observed to play a key role in suppressing star formation \citep[e.g.][]{Kauffmann2003,Hopkins2004,Croton2006,Hopkins2006,McNamara2007,Fabian2012}. X-ray, optical, and radio studies have led to the conclusion that there are at least two distinct AGN modes \citep[e.g.][]{Best2007,Hardcastle2007,Merloni2008,Best2012,Fabian2012}: a radiatively efficient mode associated with X-ray and optically-selected AGN \citep[e.g.][]{Alexander2003,Kauffmann2003,Bauer2004,Xue2010} and a radiatively inefficient mode, associated with relatively strong jets and radio emission which is common in massive elliptical galaxies with hot X-ray halos \citep[e.g.][]{Allen2006,Balmaverde2008,Dunn2010,Best2012,Pellegrini2012}. While the radiatively efficient mode appears physically related to the accretion of cold gas onto the black hole, the radiatively inefficient mode may be associated with the accretion of hot X-ray emitting gas. The kinetic energy injected into the surrounding environment by the jets emitted in this radiatively inefficient mode is, in principle, sufficient to prevent the cooling of the hot gas and suppress subsequent star formation \citep[e.g.][]{Allen2006,McNamara2007,Dunn2010,Fabian2012}.

The differences between the occurrence of AGN in cluster galaxies versus the field have not been well established. Early studies of cluster member galaxies \citep[e.g.][]{Osterbrock1960, Gisler1978, Dressler1985} determined that optically selected AGN were relatively rare in clusters. More recent work utilizing large area surveys \citep[e.g.][]{Best2007, Arnold2009, Koulouridis2010, VonderLinden2010} have expanded these findings, and prompted vigorous debate. Some studies have argued that optically luminous AGN ($\mysub{L}{[OIII]} > 10^{7} \mysub{L}{\bigodot}$) reside predominantly outside of clusters \citep{Kauffmann2004}, while others have suggested that the optical AGN activity is independent of the environment \citep{Miller2003}. Observations at radio wavelengths suggest that cluster environments are more conducive to AGN activity. Brightest Cluster Galaxies (BCG's), in particular, appear to be likely to host radio-luminous AGN while less likely to host an optically luminous AGN when compared to field galaxies of the same mass \citep[e.g.][]{Best2004, Best2007,VonderLinden2007, Dunn2010, Best2012}.  

Comparisons between the cluster and field populations of AGN using X-ray observations have led to differing conclusions, and have primarily investigated whether or not X-ray point sources are more frequently observed in the vicinity of galaxy clusters. \cite{Cappelluti2005} identify an excess of point sources in four separate cluster fields relative to expectations from field surveys, although for six other cluster fields in their study no such excess was seen. Excesses have also been observed in particular cluster fields such as 3C295, RX J003033.2+261819 \citep{Cappi2001}, and MS 1054-0321 \citep{Johnson2003}, but no excess of X-ray point sources was detected in the vicinity of galaxy clusters observed in the large field ChaMP survey \citep{Kim2004}. Other studies using larger samples of galaxy clusters have argued for a statistical excess in their fields, and have shown that the excess of sources is primarily located within the central 1-2 \Mpc \ of each cluster \citep[e.g.][]{Ruderman2005,Gilmour2009}. Multiwavelength studies of cluster galaxies incorporating optical spectroscopy along with X-ray analysis have suggested that the fraction of galaxies hosting X-ray bright AGN in clusters may increase substantially with redshift, by a factor of $\sim 8$ \citep[e.g.][]{Martini2007,Martini2009}, an effect that was also observed in the large Bo\"{o}tes field \citep{Galametz2009}. 

X-ray surveys of AGN offer higher source densities of AGN than other wavelengths \citep[see e.g.][for reviews]{Brandt2005,Brandt2010}. Deep field studies of the X-ray sky such as the \cha \ Deep Field North and \cha \ Deep Field South (hereafter \cdfn \ and \cdfs, respectively ) \citep[e.g.][]{Brandt2001, Giacconi2002, Alexander2003,Luo2008,Xue2011} have precisely measured the surface density of X-ray point sources down to fluxes of $\sim 3 \times 10^{-17} \ergpcmsqps$ in the $0.5-8.0 \keV$ band. Above fluxes of $\sim 10^{-14} \ergpcmsqps$ in the $0.5-8.0 \keV$ band, the majority of these X-ray point sources are unobscured or mildly obscured AGN \citep[e.g.][]{Bade1998,Schmidt1998,Akiyama2003}. At lower fluxes, however, other source classes such as obscured AGN, starburst galaxies, and normal galaxies become more prominent \citep[e.g.][]{Bauer2004,Lehmer2012}. Although X-ray surveys have measured the largest densities of reliably identified AGN at any wavelength, multiwavelength follow-up is necessary to test fully the impact of the cluster environment on the evolution of AGN. Field sources cannot be easily distinguished from sources intrinsic to the cluster using X-ray observations alone, even when incorporating X-ray spectroscopy. The typical spectrum for unobscured AGN in the $\sim 2-10 \keV$ \ energy band is characterized, at least to first order, by a power-law continuum, $N(E) \sim E^{-\Gamma}$, where the photon index $\Gamma$ is typically in the range of $\sim 1.5-2.5$ \citep{Nandra1994, Reeves2000, Shemmer2005,Just2007}, making it difficult to identify spectral features and so measure source redshifts. X-ray observations also provide little insight into the underlying population of galaxies in a given cluster, which diminishes the physical significance of observing a larger number of AGN in these fields. A multiwavelength study of galaxy clusters utilizing both X-ray and optical observations is well suited to offer a more complete picture of the AGN and galaxy populations within the cluster. 

Here, we present the first results of a study designed to select and observe AGN and galaxies in clusters at both X-ray and optical wavelengths. Particularly, we utilize deep observations with the \cha \ X-ray Observatory and the \sub \ optical telescope of forty-three of the most massive galaxy clusters in the redshift range $0.2 < z < 0.7$, which also have deep, multi-filter \sub \ optical imaging \citep[see][]{VonderLinden2012,Kelly2012,Applegate2012}. An important element of our work is that all of the clusters have well-determined masses based on X-ray mass proxies, allowing us to examine the AGN and galaxy properties as a function of overdensity in the cluster (i.e. relative to the virial radius as opposed to a metric aperture). In this first paper, we focus on the spatial distribution of X-ray point sources within the cluster fields. 

The structure of this paper is as follows: Section 2 discusses the sample of clusters utilized in this study. 
Section 3 discusses the initial processing of the X-ray data, while Section 4 discusses the production of X-ray point source catalogs for each cluster field. Section 5 addresses the calculation of the point-source sensitivity map for each cluster field. Section 6 presents our results on the cumulative number counts and radial profiles of X-ray point sources in these fields. In Section 7 we discuss the implications of these results on the evolution of X-ray selected AGN in galaxy clusters. For calculating distances, we assume a \LCDM \ cosmological model with \Omegam=0.3, \Omegal=0.7, and $H_{0}=70 \km \s^{-1} \Mpc^{-1}$.  

\section{The Cluster Sample}

The clusters in our study have been drawn from three wide-area cluster surveys derived from the \rosat \ All Sky Survey \citep[RASS;][]{Trumper1993}: the \rosat \ Brightest Cluster Sample \citep[BCS;][]{Ebeling1998}; the \rosat-ESO Flux-Limited X-ray Sample \citep[REFLEX;][]{Bohringer2004}; and the MAssive Cluster Survey \citep[MACS;][]{Ebeling2007,Ebeling2010}. Each sample covers a distinct volume of the Universe: the BCS covers the northern sky at $z < 0.3$; REFLEX covers the southern sky at $z < 0.3$; and MACS covers higher redshifts, $0.3 < z <0.7$, at declinations $>−40^{\circ}$. Clusters included in these samples have been instrumental in the most recent cosmological studies utilizing growth of structure tests \citep[for a review see][]{Allen2011}.  

For this paper, we restrict ourselves to the subset of these clusters with the most thorough multiwavelength follow-up measurements, including \cha \ observations  \citep{Mantz2010a,Mantz2010b} and deep multi-filter optical imaging with the \sub \ telescope. In detail, we require at least 10 ks of exposure time with \cha; robust measurements of the cluster mass and virial radius from \cite{Mantz2010a,Mantz2010b}; and that the clusters are included in the weak-lensing analysis of \cite{VonderLinden2012} \citep[see also][]{Kelly2012,Applegate2012} \footnote{As discussed in these papers, the clusters chosen for optical follow-up were selected based on their X-ray properties. This sample is therefore expected to be unbiased with respect to the optical properties of the clusters.}. In total, 43 unique galaxy clusters have been analyzed here, with redshifts ranging from  $0.2 < z < 0.7$. General information for the clusters and the \cha \ data sets used may be found in Table \ref{ChandraSample}. 

The clusters studied here are among the most massive and X-ray luminous clusters in the Universe, and host high densities of both galaxies and ICM. We therefore expect the influences of the cluster environment to be pronounced in this sample. With measurements of \rfive \ for each cluster we are able to relate observed trends in the AGN population to the virial radii of the clusters.     
Mass measurements and the associated overdensity radii, \rfive, for each cluster are taken from \cite{Mantz2010a,Mantz2010b}.  \footnote{The overdensity radius \mysub{r}{\Delta} is defined as the radius where the enclosed average mass density is equal to $\Delta$ times the critical density of the universe at the cluster's respective redshift, $\mysub{\rho}{c}(z)$. We define the virial radius for this study as \mysub{r}{200}. The corresponding mass \mysub{M}{\Delta} is defined as $\mysub{M}{\Delta} = 4/3 \pi \Delta \mysub{\rho}{c}(z)$ \mysub{r}{\Delta}$^{3}$. The mass range extends from $4 \times 10^{14} \msolar < \mysub{M}{500} < 2 \times 10^{15} \msolar$ and the overdensity radii range from $1.06  \Mpc < \mysub{r}{500} < 1.74 \Mpc$.   \
 } The typical uncertainties in measurements of \rfive \ are of order $\sim 10\%$. The \rfive \ values and X-ray centroids for the clusters are summarized in Table \ref{ChandraSample}.

\input{Table1_new}

\section{\cha \ Data Reduction}\label{ImageEmap}
All of the galaxy cluster fields were observed with the Advanced CCD Imaging Spectrometer (ACIS) aboard \cha. The majority of the observations utilize the ACIS-I chip array, but we also include observations utilizing ACIS-S. For ACIS-S observations, we only utilize chips S1, S2, and S3, for which the most accurate calibration products are available.   The standard level-1 event lists produced by the \cha \ pipeline processing were reprocessed
using the CIAO (version 4.3) software package, including the appropriate gain maps and calibration products
(CALDB version 4.4.6). Bad pixels were removed and standard grade
selections were applied. The data were cleaned to
remove periods of anomalously high background, using the standard energy ranges and binning methods recommended
by the \cha \ X-ray Center. The net exposure times after processing are summarized in Table \ref{ChandraSample}. Images for each cluster observation were produced in each of three energy bands, hereafter designated the full band ($0.5-8.0 \keV$), the soft band ($0.5-2.0 \keV$), and the hard band ($2.0-8.0 \keV$). 

For each observation, an exposure map was produced in each energy band, calculating the product of quantum efficiency and effective area across each field of view. Each exposure map was spectrally weighted over its respective energy band, assuming a photon index of $\Gamma=1.4$ \footnote{For this study, the canonical AGN source was chosen to have a photon index of $\Gamma=1.4$, to remain consistent with other studies of AGN populations in the field, in particular the \cdfs \ \citep[e.g.][]{Xue2011,Lehmer2012}.} and a hydrogen column density fixed to the value measured in the Leiden-Argentine-Bonn survey \citep[LAB,][]{Kalberla2005}. Effective exposure time maps were calculated for each observation and energy band by dividing the exposure map by its maximum effective area and multiplying by the exposure time, similar to the procedure discussed in \cite{Hornschemeier2001}. In instances where there was more than one \cha \ observation of a cluster, images and exposure maps were created for each observation individually, and then reprojected into a common aspect solution and combined into a single coadded cluster image/exposure map. These coadded images and exposure maps only include active CCD chips from the primary field of view for each observation (i.e. chips I0-I3 for ACIS-I observations, and chips S1-S3 for ACIS-S observations).

\section{Constructing the X-Ray Point Source Catalog}

\subsection{Initial Candidate X-Ray Point Sources}
An initial catalog of candidate sources was produced for each cluster using the \wav \ algorithm \citep{Freeman2002} and the full-band ($0.5-8.0 \keV$) image. Our \wav \ runs utilized a ``$\sqrt{2}$ sequence" of wavelet scales (i.e. $1, \sqrt{2}$, $2$, $2\sqrt{2}$, $4$, $4\sqrt{2}$, $8$, $8\sqrt{2}$, and $16$ pixels) and a false-positive probability threshold of $10^{-5}$. This false-positive threshold is more liberal in including faint sources than that commonly used in other studies and allows for recovering legitimate sources that fall below formally stricter thresholds \citep[e.g.][]{Alexander2001,Xue2011}. The tradeoff, however, is that we incur an appreciable number of spurious sources in our initial catalogs, many of which are associated with structures in the cluster ICM. Where all observations of a given cluster utilized a common detector (e.g. all observations were performed with ACIS-I), we utilized  \wav 's PSF information in determining the positions and size of candidate sources. The \wav \ software employs a single PSF model, and thus reliable PSF information cannot be supplied into \wav \ for the seven clusters with mixed ACIS-I and ACIS-S observations. Candidate AGN for these clusters were therefore identified assuming the smallest possible wavelet scale (1 pixel), resulting in a boosted number of candidate sources. However, our subsequent screening processes are efficient at removing spurious sources, leading to the results for these clusters being consistent with the rest. Our initial \wav \ catalogs include a total of 6266 sources across all cluster fields. Tests show that it is unnecessary to subsequently run \wav \ on the hard-band and soft-band images separately, as only a few additional sources per cluster field are identified in these images. Moreover, the majority of these additional sources detected in the hard or soft bands are subsequently rejected after further analysis.

\subsection{Improving the Initial X-ray Point Source Catalogs}\label{Refines}

Given our liberal false-positive probability threshold, incomplete incorporation of the \cha \ PSF, and the bright background ICM present at the center of every field of view, we expect that the initial catalogs produced by \wav \ have an appreciable number of spurious sources. Further analysis is required to ensure source validity. To achieve this, every candidate source in the initial catalog is re-analyzed on an observation-by-observation basis using the \ae \  point-source analysis software package
\footnote{
  The {\em ACIS Extract} software package and User's Guide are available at
  http://www.astro.psu.edu/xray/acis/acis\_analysis.html. 
  } 
\citep{Broos2010}. This offers several key improvements, specifically: (1) utilization of source and background regions that approximate the shape of the PSF calculated from ray-tracing simulations, accommodating neighboring sources and CCD chip boundaries as needed; (2) a multi-stage approach to source detection that refines the source catalog based on each source's binomial no-source probability (i.e. the probability of a source not being real given local backgrounds); and (3) better source-position determinations that improve the accuracy of each source's astrometry and photometry. We refer the reader to \cite{Broos2010} for documentation about \ae, although the most relevant aspects of the analysis are discussed below. 

Our analysis pipeline has three main stages and follows the recent study of the 4-Ms \cdfs \ \citep{Xue2011}, modified to accommodate the higher background rates and shorter exposure times of our cluster observations. Unless otherwise noted, all calculations were performed in the full band ($0.5-8.0 \keV$).

\begin{figure}
\includegraphics[width=0.92\columnwidth, angle=270]{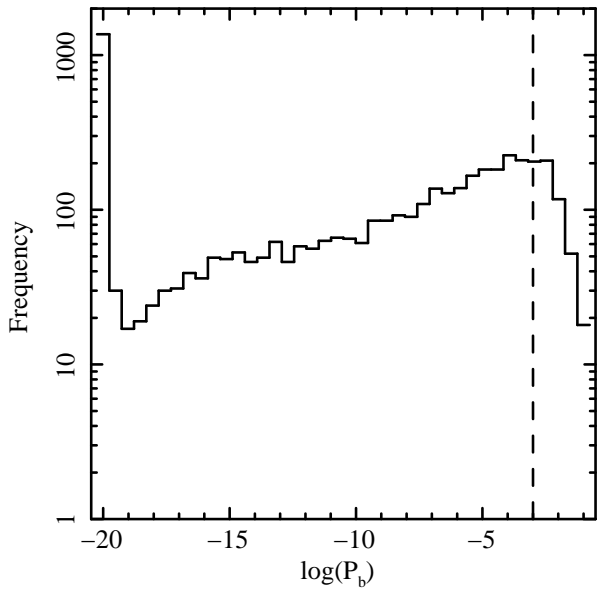}
\caption{\label{logprobs} The distribution of source detection probabilities and fluxes. (a): Distribution of X-ray no-source binomial probabilities, \prob, for all sources before inclusion in the final catalog. For this figure, probabilities were determined from counts in the full band ($0.5-8.0 \keV$). The minimum threshold chosen for inclusion in the final catalog, \thresh, is denoted by the dashed vertical line. For presentation purposes, we set $\log{\prob}=-20$ for all sources where $\log{\prob} \leq -20$.}
\end{figure}

\subsubsection{Removing False Sources}
For each observation of every candidate point source, \ae \ is used to compute a polygonal source-extraction region based on models of the \cha \ PSF at the source position. These models use simulations of the \cha \ High Resolution Mirror Assembly using the \marx \ ray-tracing simulator. Polygonal extraction regions were constructed that approximate the $\sim 90$\% encircled energy fraction (EEF) contour of the local PSF at 1.497 $\keV$. When dealing with crowded sources that have overlapping regions, \ae \ utilizes smaller extraction regions (corresponding to $\sim 40-75$\% EEFs) chosen to be as large as possible without overlapping. These source regions were used throughout the analysis, including the determination of the source's local background. For our analysis, the \ae \ ``BETTER\_BACKGROUNDS'' algorithm was utilized for background extraction, which computes the background counts within external background regions surrounding the source, accounting for contributions from the source and neighboring sources. These background regions were required to contain a minimum of 100 counts, and had a median of 111 counts. In order to provide the most accurate possible backgrounds, front-illuminated and back-illuminated chips in a particular observation were treated as separate observations. 

In order to refine the catalog and remove spurious sources, we utilize the no-source binomial probability to determine the likelihood that source counts are due to fluctuations in the background \citep[see Appendix A of][]{Weisskopf2007}. If there are a total of $S$ counts in a source region (subtending a solid angle of $\mysub{\Omega}{src}$) and \Bext \ counts in the external background region (subtending a solid angle of $\mysub{\Omega}{ext}$), the binomial probability can be calculated as:
\begin{equation}\label{Binomial}
P_{b}=\sum_{X \ge S} \frac{N!}{(N-X)!X!}p^{X}(1-p)^{N-X}
\end{equation}

\noindent In this equation, $N$ is the total number of counts in the source+background region ($S$+\Bext), and $p$= $1/(1+{\it BACKSCAL})$ is the probability that a background count is located within the source region (thus contributing to $S$), where {\it BACKSCAL}=$\mysub{\Omega}{ext}/\mysub{\Omega}{src}$. Since \ae \ calculates the expected number of background counts in the source region (\Bsrc) from this same area ratio, {\it BACKSCAL} is also equivalent to the ratio of background counts in the external background region and source regions ($\Bext/\Bsrc$). The median value of {\it BACKSCAL} for all sources included in the final catalog is roughly $62$. In the first stage of the pipeline, this probability was determined for each source and all sources with \prob $\ge 0.01$ (i.e. a probability of being a background fluctuation of more than one percent) were removed. This ensures that the source and background regions for ``real'' sources are not influenced by the presence of spurious sources nearby. Although the results vary from cluster to cluster, a total of 1407 sources (22\%)  of the initial \wav \ sources were removed by this step. Counterpart matching rates between these rejected sources and optical source catalogs are consistent with those expected solely from random coincidence, which suggests that the majority of these sources are indeed spurious and that few valid AGN are removed by this procedure. This evidence is strengthened by visual inspection of the rejected sources, which independently suggests that the vast majority of the rejected sources are indeed spurious.

\subsubsection{Improving Source Positions}\label{sourcepos}
After removal of the most likely spurious candidate sources, source and background regions were redetermined for all surviving sources. These source regions were analyzed carefully to estimate the best centroid positions. The source positions determined by \wav \ do not utilize the full information about the local PSF, in particular at large off-axis angles where the shape of the PSF becomes complex. This is especially true for sources with ``mixed'' observations (using both ACIS-I and ACIS-S), where no PSF information could be input to \wav \ in our analysis. We used \ae \ to improve the source positions, utilizing three different algorithms: (1) calculating the centroid from the data ({\it DATA});  (2) correlating the PSF within the neighboring region ({\it CORR}); and (3) reconstructing an image of the region neighboring the source using the background and source information and a maximum likelihood algorithm ({\it ML}). Tests carried out by the \ae \ team \citep{Broos2010} provide guidance as to which of these algorithms is best suited for a particular source, based on its position and the positions of neighboring sources. Following their recommendations, we have moved all sources to their {\it DATA} positions when they were uncrowded and located near the aimpoint (within 8\arcmin); to their {\it CORR} positions when the sources were far from the aimpoint; and to their {\it ML} positions whenever other sources are sufficiently close that the wings of their respective PSF's may overlap. The typical displacement from the original \wav \ positions is $\lesssim 1\arcsec$.

\begin{figure}
\includegraphics[width=0.92\columnwidth, angle=0]{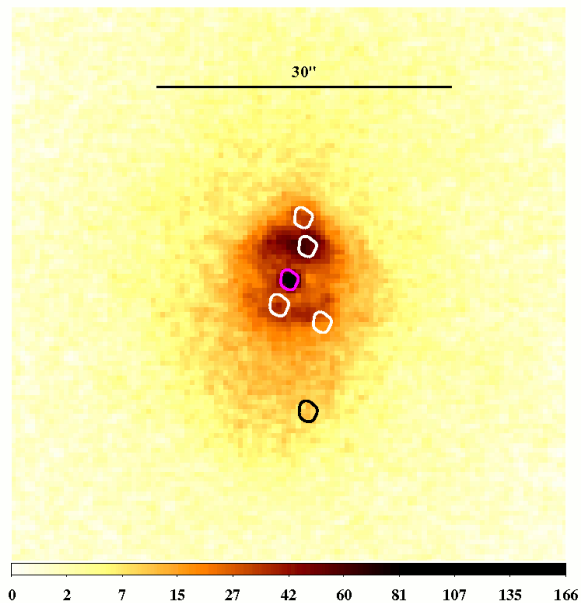}
\caption{\label{centersources} An example of spurious candidate sources initially identified within the central $\sim 1 \arcmin$ of \MACS. Contours show the respective PSF region ($\sim 90\%$ EEF) for each candidate source. This particular cluster shows evidence for both a recent merger and powerful AGN feedback \citep{Ehlert2011}, which results in the sharp fronts and cavities in the $\sim 50 \kpc$ surrounding the cluster center. The black source has been identified as spurious automatically by \ae \ (\unthresh ), while the white sources had to be removed manually. Although these sources are significant compared to their local background, they are not AGN. Such candidate sources can only be reliably removed by visual inspection. The only true AGN source included in the final catalog is the magenta source at the center. This is one of only two luminous cD galaxies in this sample ($\mysub{L}{X} \sim 8 \times 10^{43} \ergps$). }
\end{figure}

\subsubsection{Source Photometry and Spectroscopy}\label{photometry}
With the source positions set, final source products were extracted from each observation. For each new source position, the PSF was simulated by \ae \ at the nominal PSF energy (1.497 \keV) as well as four other energies (0.277, 4.510, 6.400, and 8.600 \ in order to determine the best estimate of the PSF. Background regions and models were recalculated for every source, and spectral products including ancillary response files and response matrices were extracted, based on the model PSF region. The spectral products, PSF models, and no-source probabilities were combined for sources observed multiple times. In order to determine source fluxes, we utilized the photon flux measurements for each source computed by \ae \footnote{We have used both the ``FLUX1'' and ``FLUX2'' measurements made by \ae. The ``FLUX1'' measurement is defined as the net counts in each channel divided by the ACIS ancillary response in the channel (for that source position in the same energy band) and the exposure time, summed over all channels. The ``FLUX2'' measurement is defined as the net counts in an energy band divided by the mean ancillary response of the ACIS detector and the exposure time. Both of these measurements convert the measured counts into units of photons $\cm^{-2} \s^{-1}$.} and converted them to energy fluxes by assuming a power-law model with photon index $\Gamma=1.4$. This value is consistent with the expected photon index for the field population of AGN. Spectral analyses of the individual sources in our catalog are consistent with this assumption. Statistical uncertainties on the source fluxes are typically at the $\sim 30\%$ level, taking into account uncertainties in the source count rates, background, and best-fit power-law model parameters. For sources with successful spectral fits, the median photon index is $\Gamma=1.43$, while the mean is $\Gamma=1.37.$

\subsubsection{Final Source Selection}

Our threshold for inclusion in the final point source catalog is \thresh \ in any band. These thresholds were chosen to be conservative. As a comparison, the \cdfs \ uses a threshold of $\prob < 0.004$ in any band and led to a sample of high purity (i.e. a catalog with few spurious detections expected based on simulations and optical counterpart matching rates). A total of 4210 sources are included in the final catalog. The distribution of \prob \ in the full band for all sources is shown in Figure \ref{logprobs}.

Using estimates of the PSF and local background, we are able to efficiently remove the majority of spurious sources identified by \wav, including sources associated with diffuse, extended ICM emission. However, we also investigated the central regions of every galaxy cluster visually. Despite the best attempts of \ae \ to suppress sources associated with the diffuse ICM, structures that vary significantly on angular scales comparable to the \cha \ PSF can be mistaken for point sources at high formal significance, and can only be reliably removed with visual inspection. The vast majority of these sources are located within $\sim 100 \kpc$ of the cluster centroids, and are predominantly located in clusters identified as hosting cool cores \citep[e.g.][]{Dunn2008,Hlavacek2012}. Cool core clusters typically have sharply peaked surface brightness profiles \citep[e.g.][]{Peterson2006}. The presence of a cool core is also closely linked to AGN feedback by the cluster's central galaxy \citep[e.g.][]{Fabian2003,Fabian2006,McNamara2007,Sanders2007,Million2010,Werner2010}, and ICM substructures arising from this feedback such as shock fronts or cavities can also result in structures falsely classified as bright point sources. Since the cool cores are typically associated with the centers of clusters where the galaxy density is also highest, these sources also commonly have optical counterparts by random coincidence. Examples of such false candidate sources are shown in Figure \ref{centersources} in the extreme cool core cluster \MACS \ \citep{Ehlert2011}. In our visual inspections, we were conservative, removing all candidate sources that could not be clearly identified as bright AGN visually. A total of $\sim 20$ sources were removed in this fashion. It is possible, therefore, that our results slightly underestimate the AGN density in the innermost regions of the clusters ($r \lesssim 50 \kpc$). Since the cluster emission is brightest and most structured near the cluster center, the properties of all point sources identified in these regions will be subject to systematic uncertainties due to potential errors in the background estimation. Based on our conservative visual selection, only two X-ray luminous AGN are unambiguously identified within 50 \kpc \ of the cluster centroids: one in MACSJ1427.2+4407, and the other in \MACS. Additionally, these visual inspections enabled us to manually remove sources that were correlated with instrumental artifacts unrelated to physical point sources.

\subsubsection{Initial Optical Counterpart Matching}\label{opticalcp}
Although we reserve detailed counterpart matching for future work, we have performed an initial check comparing the fraction of matches between our X-ray point sources and catalogs of optical sources derived from \sub \ imaging in hand \citep{VonderLinden2012,Kelly2012,Applegate2012}. The optical catalogs include all source types, and are complete to a magnitude limit of $\sim 25$ in the \sub \ $R$-band. We adopted a constant matching radius of $2\arcsec$ between X-ray and optical sources, and in instances where more than one candidate optical counterpart within $2\arcsec$ of the X-ray source was present, the brightest optical source was chosen as the counterpart. \footnote{We find that a $2\arcsec$ matching radius maximizes the counterpart matching rate without introducing a significant number of coincidence matches, and is sufficiently large to account for any systematic pointing offsets that may exist between the \cha \ and \sub \ data.}  In total, the number of sources with an optical counterpart brighter than magnitude 25 (\sub \ R-band) is 2826 out of 4210, or $67\%$. 164 sources in the X-ray catalogs reside in regions of the optical images unsuitable for reliable photometry (i.e near CCD chip gaps or bright stars), so no counterparts for those sources could be established. The counterpart matching rate between X-ray sources and R-band sources is slightly lower than the \cdfs \ at the same flux and magnitude limits \citep[e.g.][]{Xue2011}, as can be expected given the greater depth and wavelength coverage of the follow-up data used for that survey. \footnote{As an example, for sources brighter than $3 \times 10^{-15} \ergpcmsqps$ in the soft band, the fraction of X-ray sources with optical counterparts with $R < 25$ is $\sim 91\%$ (31 out of 34), while $\sim 86\%$ of the X-ray point sources in this sample at the same flux limit have optical counterparts with $R < 25$. } For sources with successful optical counterparts, the median separation between the X-ray source and its optical counterpart is approximately $0.7\arcsec$.

\subsubsection{Source Extension}

We have performed a check for spatial extension of our sources using the Kolmogorov-Smirnov (K-S) test discussed by \cite{Xue2011}. For each detected X-ray source we derived cummulative EEFs for both the PSF model and source counts out to 90\% EEF. We then used a K-S test to compute the probability ($\mysub{P}{KS}$) that the two sets of cummulative EEFs are consistent with one another. A total of 236 sources out of 4210 ($5.7\%$) have a measured value of $\mysub{P}{KS} < 0.05$ (i.e. have a distribution of source counts inconsistent with the simulated PSF at $\geq 95\%$ confidence), and a total of 127 sources ($3\%$) have $\mysub{P}{KS} < 0.01$. These results are consistent with measurements from the \cdfs \ \citep{Xue2011}. We conclude that at most $\sim 2\%$ of the sources may have spatial extension. Visual inspection of the sources with $\mysub{P}{KS} < 0.01$, however, suggests that the majority of these sources do not have significant extension beyond the PSF, especially for sources observed multiple times at different off-axis angles. We therefore treat these results as a conservative upper limit to the true number of extended sources in our catalogs.

\subsection{The Final Point Source Catalog}\label{catalogsec}

\begin{figure*}
\includegraphics[width=\textwidth]{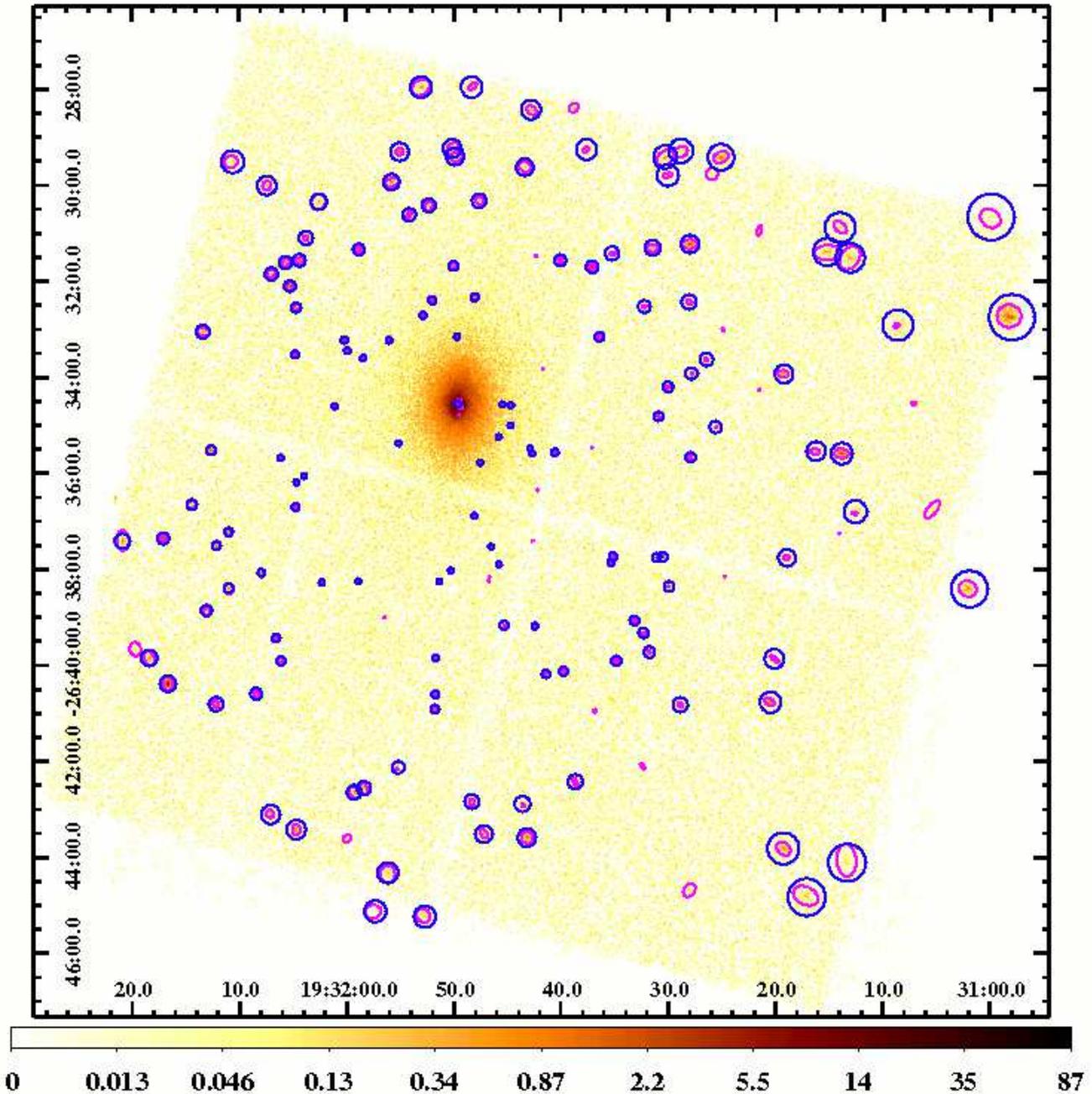}
\caption{\label{pointsourceimage} The \cha \ full-band counts image for the galaxy cluster \MACS \ \citep{Ehlert2011}, with the initial and final point source catalogs overlaid. The magenta ellipses correspond to the initial point sources detected by \wav, while the blue circles correspond to point sources that satisfy the threshold \thresh \ in the full band. The radius of the blue circle corresponds to 1.5 times the 99\% EEF radius for the point source position. For this cluster, 162 initial candidate sources were detected by \wav. This was refined to a final sample of 137 high-confidence (\thresh) sources detected across all three bands.}
\end{figure*}

The final point-source catalog has a total of 4210 X-ray bright point sources detected at our selection criteria (\thresh \ in any band) across a total survey area of 4.2 $\deg^{2}$. In the full ($0.5-8.0 \keV$), soft ($0.5-2.0 \keV$), and hard ($2.0-8.0 \keV$) bands, a total of 4099, 3344, and 2629 sources, respectively, satisfy the selection criteria for that specific band. The majority of these sources (3732) are detected in at least two bands, and 2019 sources were detected in all three bands. Only 256, 81, and 30 sources satisfy the selection criteria exclusively in the full band, soft band, and hard band, respectively. Treating the no-source binomial probability formally, there is a $\lesssim 5\%$ chance of there being more than seven spurious sources in any band. \footnote{Since the no-source binomial probability is a formally derived null-hypothesis test \citep{Weisskopf2007}, the true false-positive rate should not deviate significantly from these estimates. This is slightly complicated by the possibility of extended sources at high significance, but nevertheless we expect that these catalogs are not contaminated by large numbers of spurious sources. } General information about the point source catalogs for each cluster can be found in Table \ref{ChandraProperties}. As an example, the initial and final point source positions for the galaxy cluster \MACS \ are shown in Figure \ref{pointsourceimage}. 

\section{Sensitivity Maps}
The minimum flux at which point sources can be reliably detected in a given \cha \ observation and energy band depends on the exposure time of the observation, the local background, vignetting corrections, the local PSF, the assumed properties of the source, and the manner in which point sources are selected. For each cluster field, we calculate the flux limit at every position in the field of view for a canonical AGN source, using the effective exposure maps (\S \ref{ImageEmap}), background maps (\S \ref{BGsect}), and the no-source binomial probability (Equation \ref{Binomial}). Sensitivity maps were created in all three science bands.

\subsection{Background Map Creation}\label{BGsect}
Background maps for each cluster and energy band were created directly from the cluster data,  masking all point sources in the final catalogs. Masked regions were re-filled using the {\it CIAO} routine {\it dmfilth}, which replaces the excised regions with random values drawn from a Poisson distribution, the mean of which is sampled from a unique background region for each source. For the excised source regions, we used circles with radii 1.5 times the $ 99\%$ enclosed energy fraction radii for each source. For refilling, we used annuli with inner and outer radii of 1.5 and 2.5 times the $\sim 99\%$ enclosed energy fraction radii. Approximately $\sim 25 \%$ of the pixels in each image were re-filled by this procedure. The resulting background map is the sum of unresolved cosmic sources, diffuse cluster emission, Galactic foreground emission, and instrumental background from each observation. In this study, resolving the relative contributions of these background components is not important, as the sensitivity to point sources only depends on the total background present in each pixel.

\subsection{Deriving the Sensitivity Maps}

With the total background maps in hand, we determined the flux limit for point sources anywhere in a given cluster field. We determine the minimum number of counts required for source detection by solving Equation \ref{Binomial} for our catalog threshold no-source binomial probability, \thresh. To do so, we require estimates of the background counts in both a source aperture (\Bsrc) and an external background region (\mysub{B}{ext}), both of which will depend strongly on the local PSF. For these calculations, the PSF was assumed to be a circular aperture with a radius that depends only on the angular distance between the source position and the exposure-weighted aimpoint, \mysub{\alpha}{p}. The radius of the aperture was estimated using the $\sim 90\%$ EEF radius of point sources at the same off-axis angle. Determinations of the background counts in the source aperture, \Bsrc, are taken directly from the background maps discussed above, using the local PSF aperture. Estimates of the external background counts, \Bext, are determined from the \Bext \ values used in the cluster point source catalog. We set the \Bext \ to the maximum \Bext \ value of the cluster-catalog sources that are located in the same annular region as the position in question, with the inner/outer radii being $\alpha_{p}-1.0$\arcmin and $\alpha_{p}+1.0$\arcmin. By choosing the maximum value of \Bext, the ratio {\it BACKSCAL} is maximized for every source aperture, meaning that the counts required for source detection in an aperture is minimized.   

\begin{figure*}
\includegraphics[width=\textwidth]{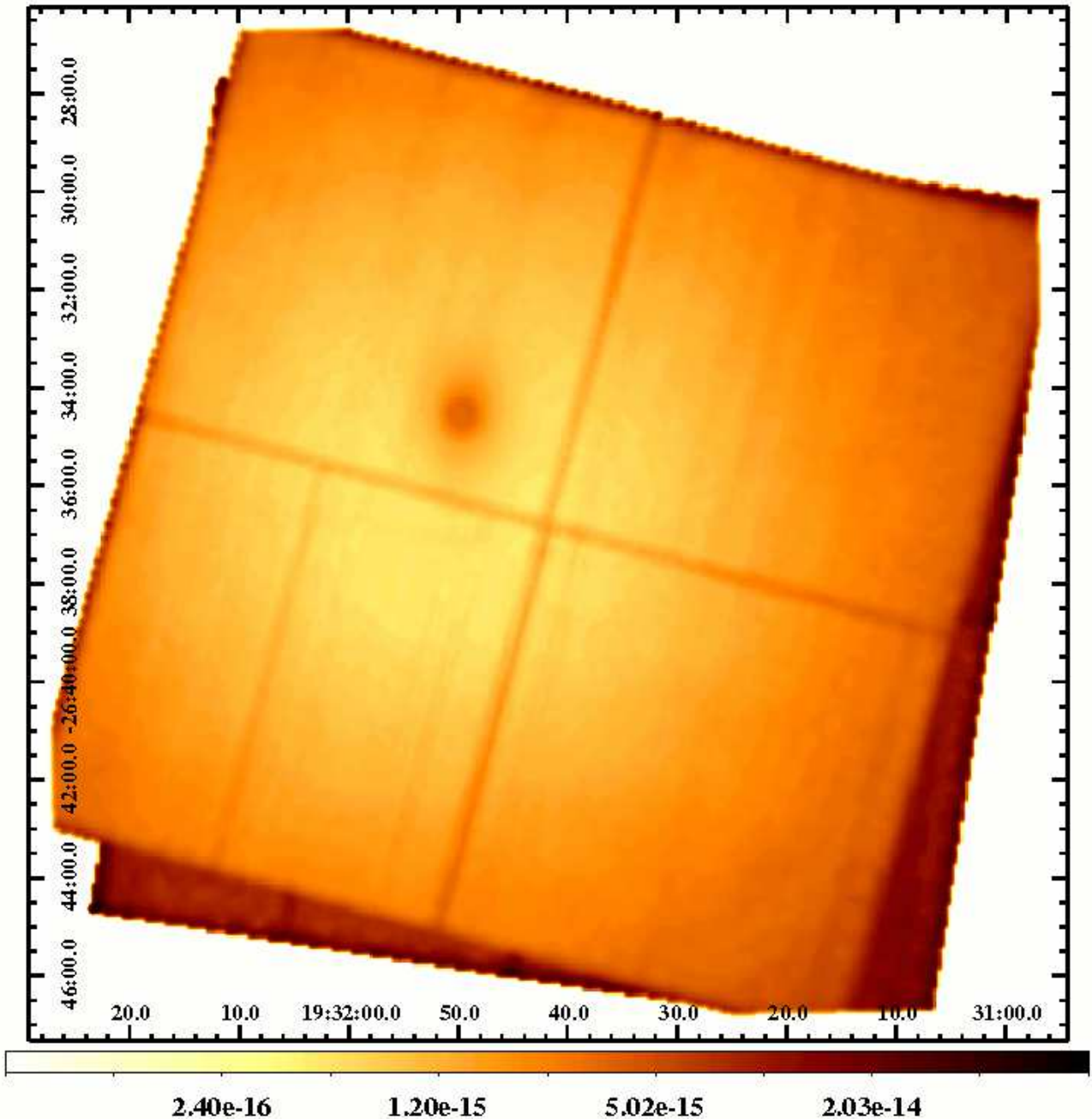}
\caption{\label{senmap} The sensitivity map for \MACS \ in the 0.5-8.0 \keV \ energy band, calculated by solving for the number of counts required to detect a source at a binomial probability of \prob= $10^{-3}$. These count estimates were converted to a physical flux assuming an absorbed power-law spectrum with a photon index of $\Gamma=1.4$. For this particular cluster, the minimum full-band flux to which 50\% of the area is sensitive is $1.8 \times 10^{-15} \ergpcmsqps$. The dark region at the center of the cluster shows that the local flux limit near the cluster is approximately 5-10 times higher than the surrounding region despite being the region with the sharpest PSF, due to the presence of the bright cluster ``background'' emission.   }
\end{figure*}

With estimates of \Bsrc \ and \Bext \ for each position in the background map, we numerically solve the binomial probability equation to determine the minimum number of counts required for detection under the source-detection criterion. These counts are then converted into a count rate given the effective exposure map (\S \ref{ImageEmap}), and the count rate is converted into a physical flux assuming an absorbed power-law spectrum with a photon index of $\Gamma=1.4$. The above procedure takes into account PSF broadening with off-axis angle, the variations in the effective exposure time (due to vignetting and CCD chip gaps, for example), and variations in the background across the field of view, including the diffuse galaxy cluster emission. \footnote{Only one sensitivity map is determined for each cluster, as this procedure is equally valid for clusters observed multiple times. The only difference is that the background maps and exposure maps from each individual exposure are summed together before the local flux limit is calculated. The median distance between any observation aimpoint and the mean aimpoint is $\sim 0.7\arcmin$, therefore changes in the local PSF between observations should not be significant. For a few clusters, observations have aimpoints appreciably far away from the mean aimpoint ($\gtrsim 2\arcmin$), which adds a source of systematic uncertainty to the sensitivity maps. These systematic uncertainties are not large compared to the total survey area over all fields, and are therefore inconsequential for the final results.  } An example sensitivity map, for the cluster \MACS, is shown in Figure \ref{senmap}. The impact of the diffuse cluster emission on the local flux limit is apparent at the center of the sensitivity map, where the local flux limit is a factor of $\sim 5-10$ times higher at the cluster center compared to the surrounding region, despite the significantly sharper PSF at the cluster center. We define the band-specific flux limit for each cluster field as the minimum flux in that energy band to which 50\% of that particular field is sensitive. These flux limits range from $1-9 \times 10^{-15} \ergpcmsqps$ in the full band, and are shown for each cluster and energy band in Table \ref{ChandraProperties}.

\input{Table2_new}

\section{Results}
Our final source catalogs and sensitivity maps allow us to determine the distribution of X-ray bright point sources across the cluster fields. We present two main results in this study: the cumulative number counts of point sources per unit sky area across the fields; and their radial distributions. In both cases, the dominant uncertainty is the Poisson uncertainties in the total number of sources. The expected Poisson fluctuations for a sample of size $n$ is estimated using the 1-$\sigma$ asymmetric confidence limits of \cite{Gehrels1986}.  The 1-$\sigma$ upper confidence limit $\lambda_{U}$ and 1-$\sigma$ lower confidence limit $\lambda_{L}$ for a sample of $n$ sources are approximated as  

\begin{eqnarray*}
\lambda_{U}(n)=n + 1 + \sqrt{0.75 +n} \\ 
\lambda_{L}(n)=n\left(1-\frac{1}{9n}-\frac{1}{3\sqrt{n}}\right)^{3}
\end{eqnarray*}

\noindent which are both accurate within a few percent for all values of $n$. These Poisson uncertainties, which are at the $\sim 5\%$ level, dominate over uncertainties in the source flux measurements and uncertainties in the sensitivity maps. \footnote{This has been confirmed by Monte Carlo simulations, calculating the logN-logS number counts from random realizations of the source fluxes and sensitivity maps. The uncertainties on the source fluxes were taken from our spectral fits, while the uncertainties in the sensitivity maps were assumed to be $\sim 20\%$ and applied coherently across an entire field of view. Both effects were shown to contribute negligibly to the overall uncertainty when compared to the Poisson fluctuations on the detected sources. } Our results do not strongly depend on the choice of photon index for the canonical AGN source. In all instances, we compare our results to the results from both the \cdfs \  and the \cha \ {\it COSMOS} survey \citep[e.g.][]{Elvis2009,Puccetti2009} in the same energy band. These surveys should provide estimates of the number density of field sources not associated with the clusters, at least within the uncertainties associated with cosmic variance \citep[e.g.][]{Lehmer2012}. We note that the {\it COSMOS} results presented here are obtained from re-analyzing the \cha \ {\it COSMOS} field data using our pipeline. This ensures that discrepancies between the clusters and field surveys due to differences in the analysis procedure and calibration are minimal. We omit comparisons to published results based on the \xmm \ {\it COSMOS} survey \citep[e.g.][]{Cappelluti2009} due to systematic differences in flux calibration, which are estimated at the $\sim 10\%$ level in the soft band \citep{Nevalainen2010,Tsujimoto2011}. Such systematic uncertainties in the source fluxes lead to systematic uncertainties in the cummulative number counts at a given flux of approximately $\sim 30\%$. Additionally, \cha \ and \xmm \ calibration differences are energy dependent, and will affect the hard and soft bands differently. 

\begin{figure}
\includegraphics[width=0.92\columnwidth, angle=270]{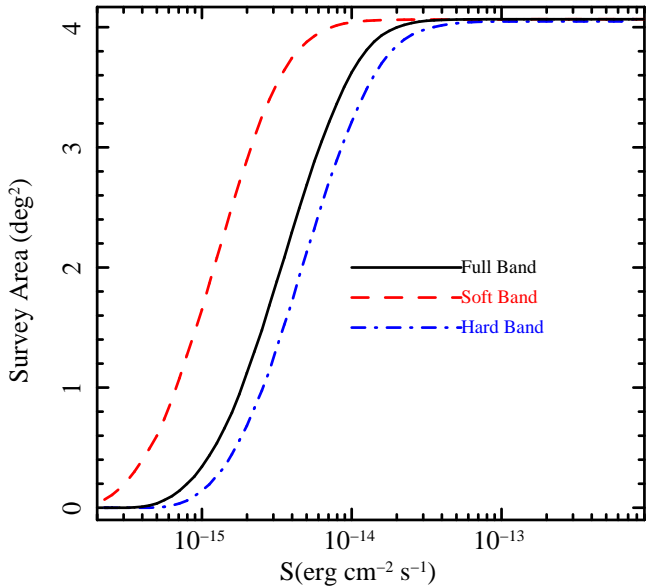}
\caption{\label{SurveyArea} Survey solid angle as a function of flux limit in the soft, hard, and full bands for all cluster observations. In order to ensure survey completeness, we have only included the region within 15 arcminutes of the aimpoints in calculating the survey area presented here. The total survey area within 15 arcminutes of the aimpoints over these 43 clusters is $4.06 \deg^{2}$.}
\end{figure}

In order to ensure survey completeness, we restrict ourselves to the survey area within a 15 arcminute radius of the aimpoints in calculating the logN-logS cumulative number counts and radial profiles. As we will show, we can reliably measure the projected source density within $\sim 3$ \rfive \ or 3.5 \Mpc \ of the clusters after imposing this cut.  
The total survey area sensitive to different flux levels after imposing this 15 arcminute cut is shown in Figure \ref{SurveyArea}. This shows a steep drop in survey area at fluxes below $\sim 3 \times 10^{-15} \ergpcmsqps$ ($0.5-8.0 \keV$). There is comparable sensitivity to a given flux in the hard and full bands, due to relaxed selection criteria in the hard band.

Only a small number of sources in the final catalog within 15 arcminutes of the aimpoints have measured full-band fluxes below the flux limit at their respective positions ($282/4084 \sim 6.9\%$), and, in the majority of these cases the flux measurements are consistent with the flux limits within statistical uncertainties. Only 73 (1.8\%) of the point sources in the final catalog have a measured flux inconsistent with the local flux limit at their respective positions at a level greater than $68\%$ confidence at their respective positions. It is expected that some sources will have such characteristics, given differences between the spectra of the point sources and the canonical AGN spectrum ($\Gamma=1.4$). The sources with fluxes below their local sensitivity limits have spectral fits that are significantly softer than the canonical $\Gamma=1.4$, with a median value of $\Gamma \sim 2$.

\subsection{Cumulative Number Counts}

\begin{figure*}
\centering
\subfigure[]{
\includegraphics[width=1.3\columnwidth, angle=270]{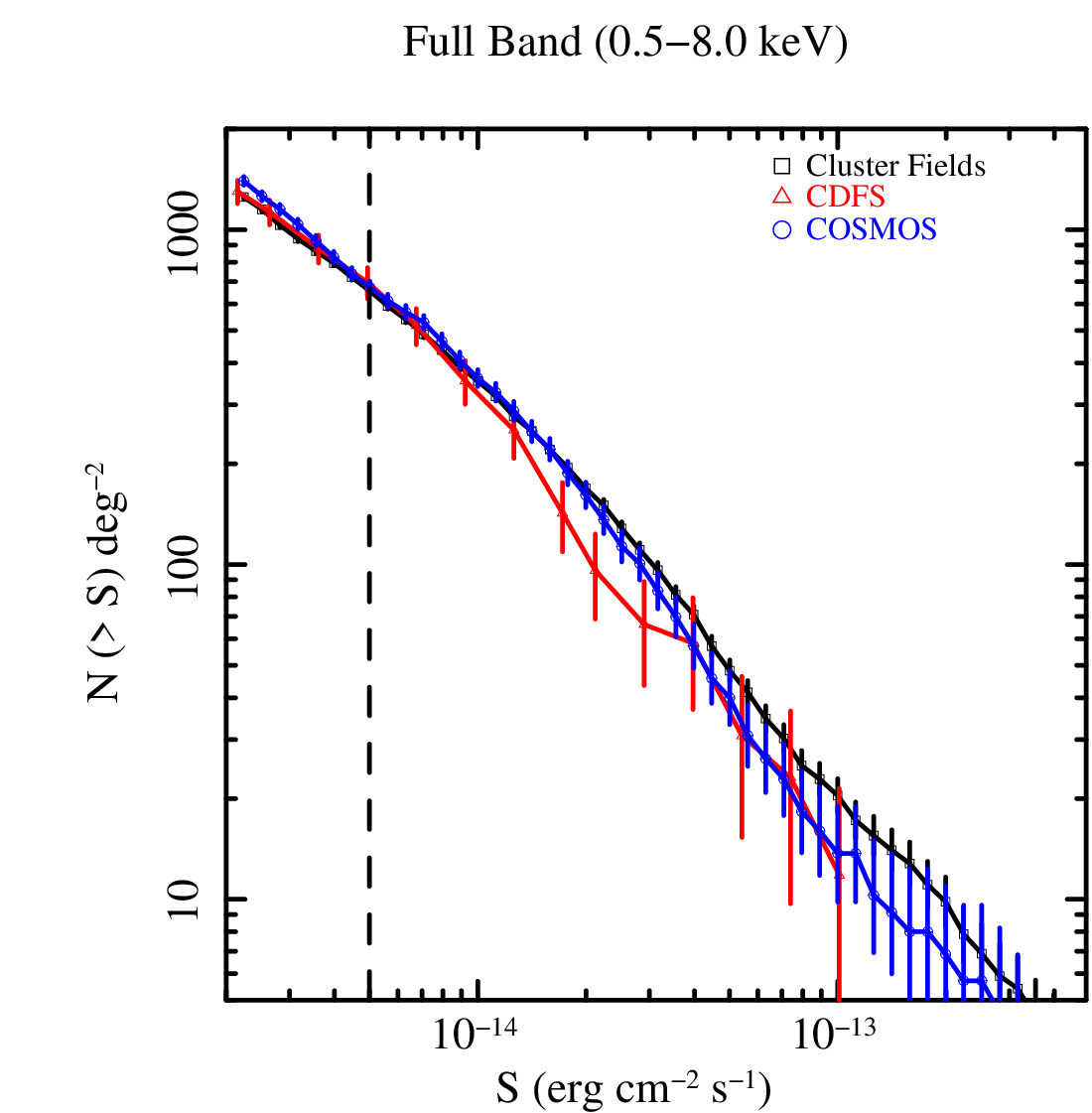}
\label{FulllogNlogS}
}
\subfigure[]{
\includegraphics[width=0.9\columnwidth, angle=270]{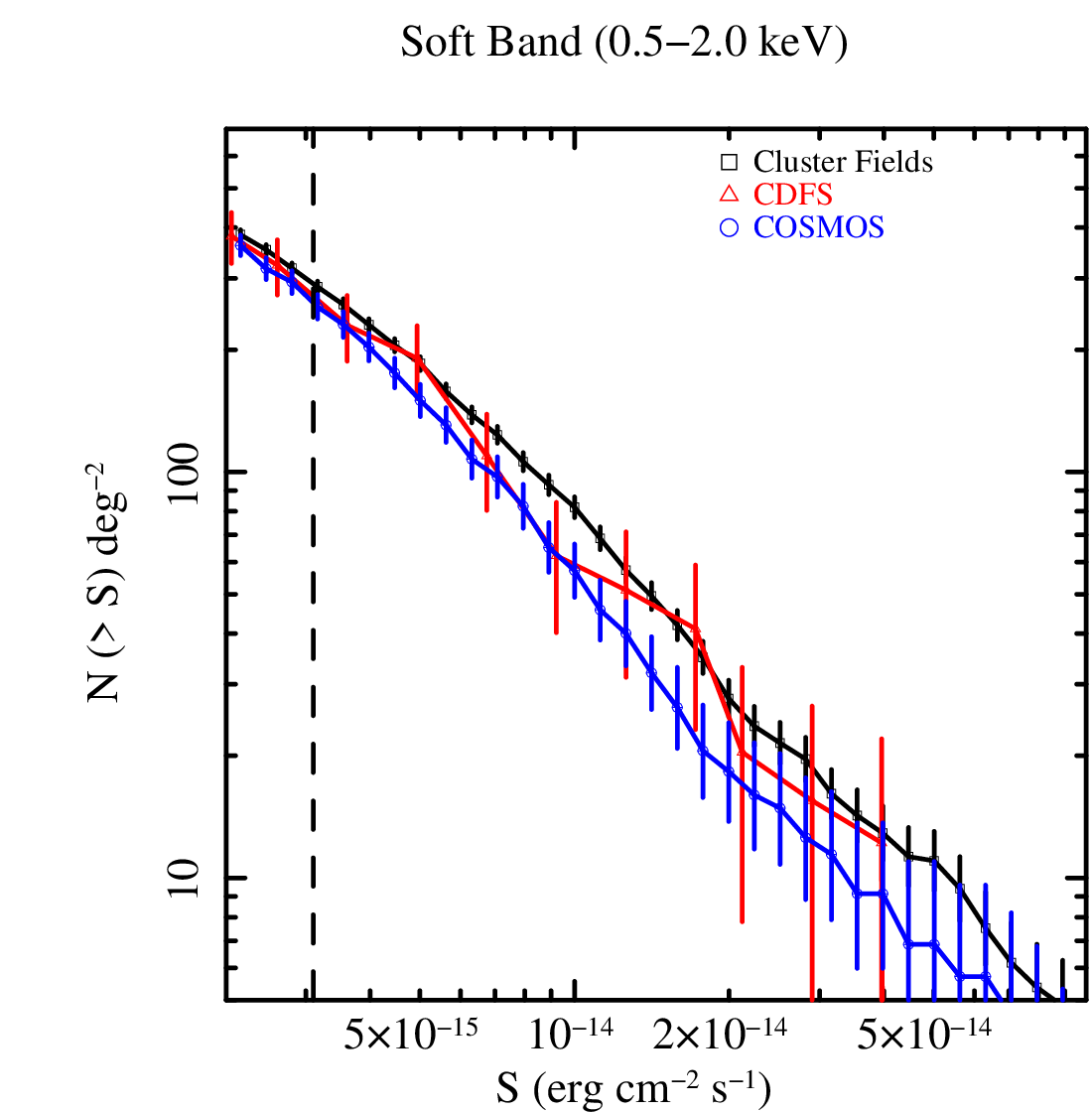}
\label{SoftlogNlogS}
}
\subfigure[]{
\includegraphics[width=0.9\columnwidth, angle=270]{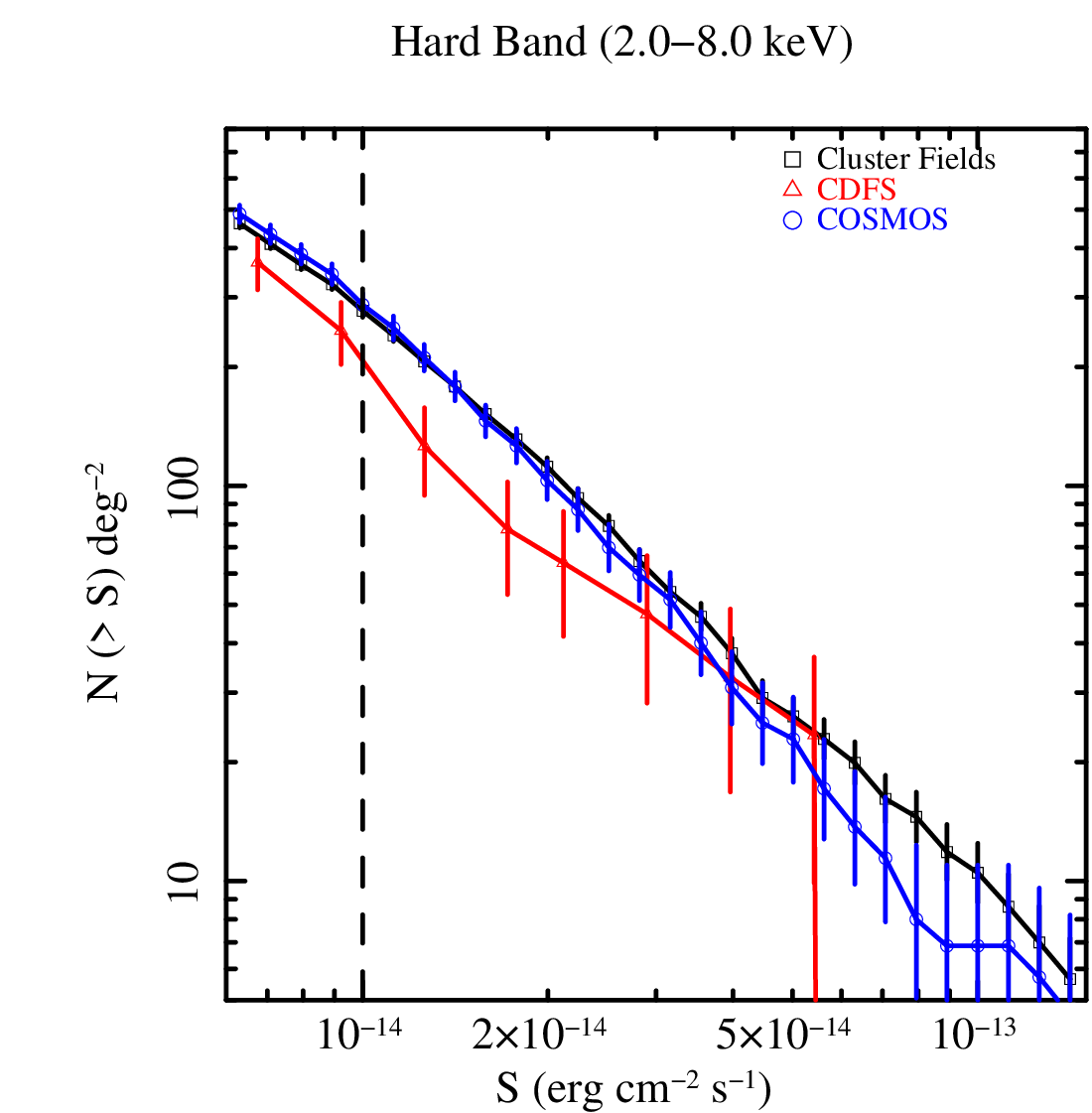}
\label{HardlogNlogS}
}

\caption{\label{NumberCounts} Cumulative number counts (logN-logS) in all three energy bands for the cluster field X-ray point source catalog in black. In red are the cumulative number counts for the \cdfs \ in the same energy band \citep{Lehmer2012}, and in blue are the results of our re-analysis of the \cha \ {\it COSMOS} field \citep{Elvis2009}. We only include bright sources in determining the radial distribution of AGN, the band-specific flux limit of which is denoted by the vertical dashed line in each figure. (a): The full-band cumulative number counts. (b): The soft-band cumulative number counts. (c): The hard-band cumulative number counts.  All three of these logN-logS curves show similar trends, with a small excess of sources in the cluster fields detected at high fluxes.}

\end{figure*}

The cumulative number density of sources above a given flux ($S$) is calculated as 

\begin{equation}
N (> S) = \sum_{S_{i} > S} \frac{1}{\Omega_{i}}
\end{equation}

\noindent where $\Omega_{i}$ is the total survey area sensitive to the $i^{th}$ source flux $S_{i}$. The $\log{N}-\log{S}$ cumulative number counts for sources in the soft, hard, and full energy bands are shown in Figure \ref{NumberCounts}, together with the \cdfs \ and {\it COSMOS} results in the same energy bands \citep[][]{Lehmer2012}. All fluxes have been corrected for Galactic absorption. The cumulative number counts in all three bands for the cluster fields, the \cdfs, and {\it COSMOS} show similar shapes, consistent with the frequently observed broken power-law shape  \citep[e.g.][]{Cowie2002,Moretti2003,Bauer2004,Lehmer2012}. A small excess of sources is observed at higher fluxes in all three energy bands, and at low fluxes all three bands are consistent with the expected field population, although the largest relative excess is observed in the soft band.

\subsection{The Radial Distribution of X-ray Sources}\label{RadProfSec}
\begin{figure*}
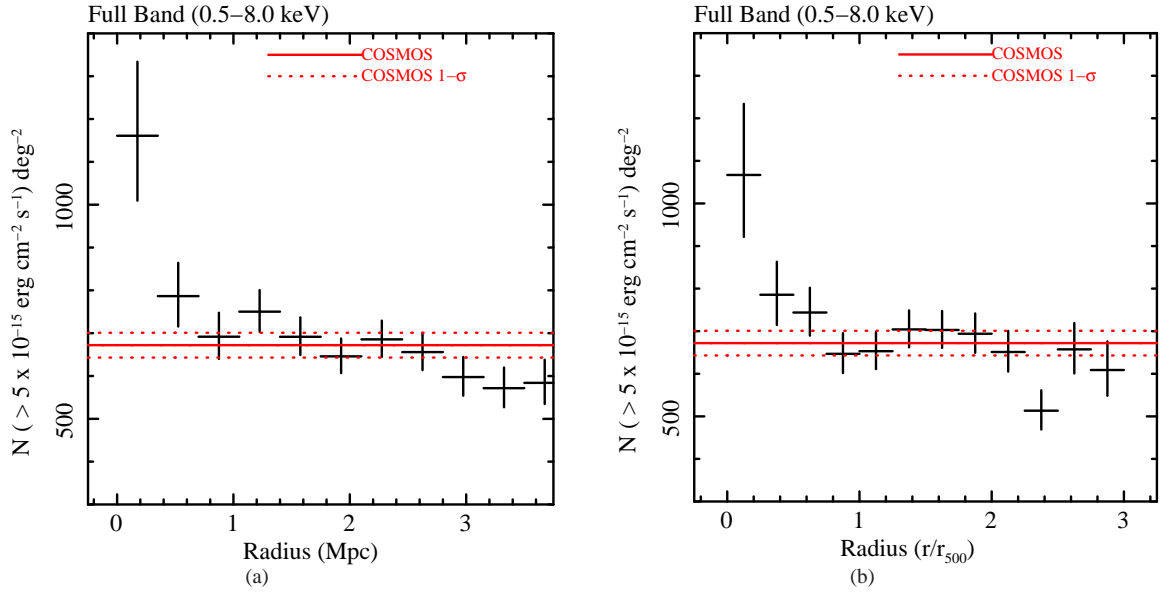

\centering
\subfigure[]{
\includegraphics[width=3in, angle=270]{MpcProf_lowflux.ps}
\label{MPCplot}
}
\subfigure[]{
\includegraphics[width=3in, angle=270]{R500_lowflux.ps}
\label{R500plot}
}
\caption{\label{RADplots} The projected density of X-ray point sources in the full band ($\mysub{F}{X}(0.5-8.0 \keV) > 5 \times 10^{-15} \ergpcmsqps$), in units of $\rm{deg}^{-2}$. For both figures, the solid red line denotes the best fit source density of {\it COSMOS} \ at the corresponding flux limit, while the dashed red lines correspond to its $1-\sigma$ confidence limit. (a): The projected density of sources as a function of radius in units of \Mpc, assuming each point source is at the same redshift as the cluster. A total of 2056 sources were included in the calculation of this profile. A large excess is seen in the central $\sim 250 \kpc$ of the cluster centers. (b): The same surface density as (a), instead normalized to each cluster's characteristic radius \rfive. In these units the central density peak is more broadly distributed throughout the central $\sim 0.5$ \rfive \ of the clusters. In both cases, the measured density of background sources is consistent with expectations from {\it COSMOS}. {\bf The black curve shows the most probable projected source density profile from our MCMC analysis; a power law with index $\beta=-0.7$. The background source density for this model is slightly lower than but statistically consistent with {\it COSMOS}.}   }

\end{figure*}

\begin{figure*}
\centering

\subfigure[]{
\includegraphics[width=3in, angle=270]{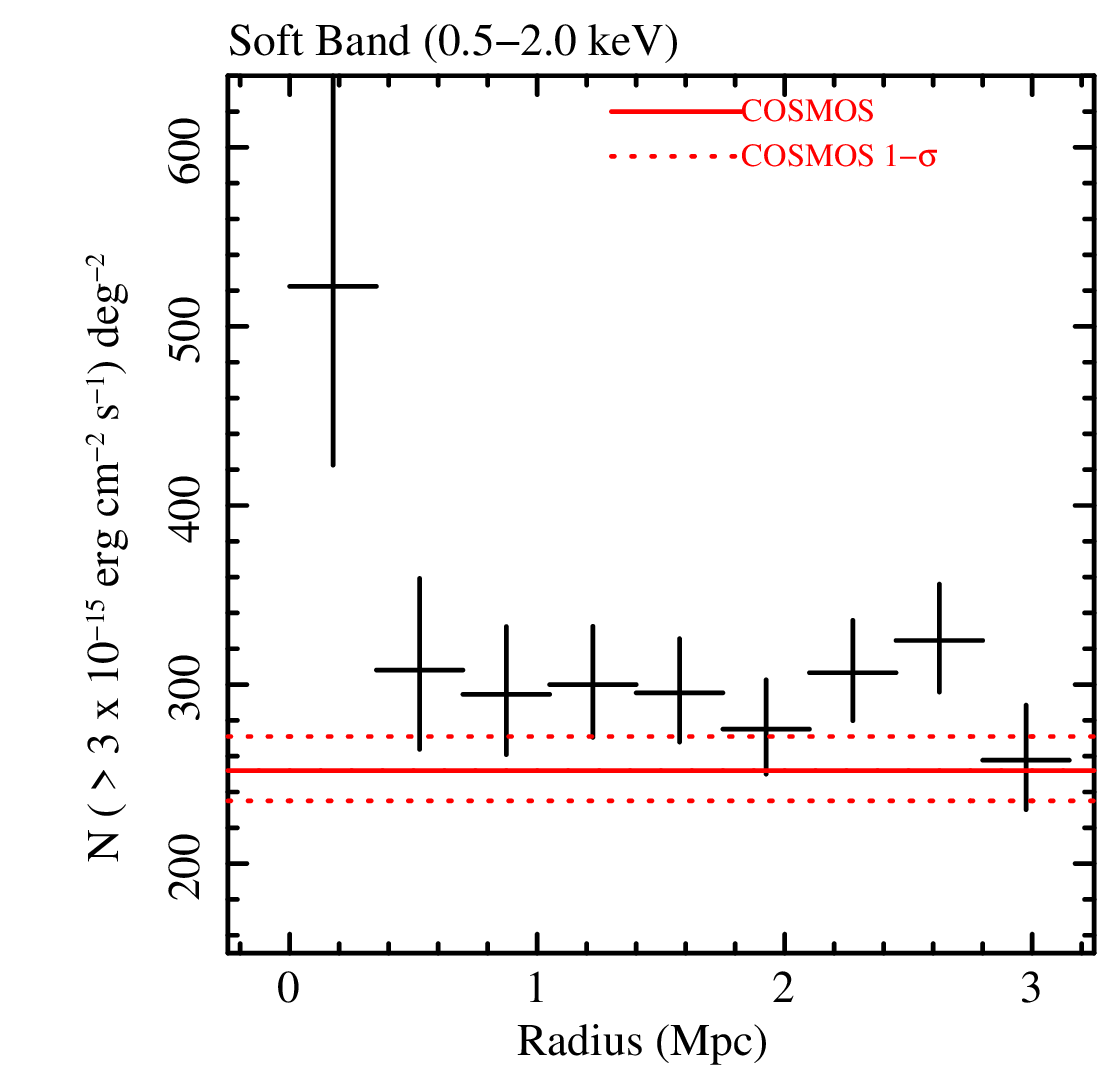}
\label{Mpcsoftplot}
}
\subfigure[]{
\includegraphics[width=3in, angle=270]{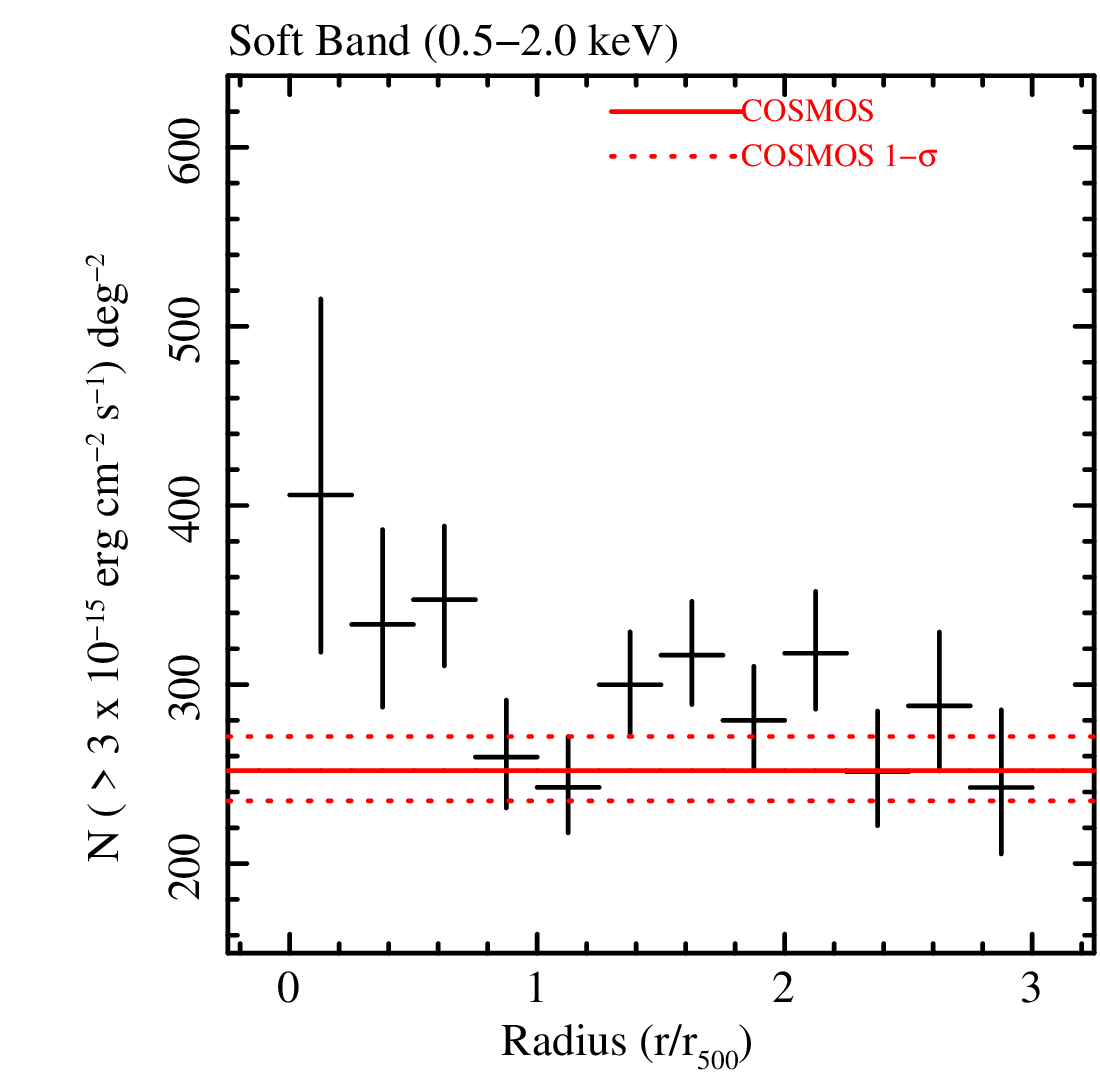}
\label{R500softplot}
}

\caption{\label{Radsoftplots} The projected density of X-ray bright point sources in the soft band ($\mysub{F}{X}(0.5-2.0 \keV) > 3 \times 10^{-15} \ergpcmsqps$). For both figures, the solid red line denotes the best fit source density of {\it COSMOS} at the corresponding flux limit, while the dashed red lines correspond to its $1-\sigma$ confidence limit. (a): The surface density of X-ray bright soft band sources as a function of radius, in units of \Mpc. A total of 968 sources were included in the calculation of this profile. (b): The surface density of X-ray bright soft band sources as a function of radius, in units of \rfive. Just as in the full band, using the overdensity radius \rfive \ results in a broader central excess of sources. In both profiles shown here, there is evidence for a second excess near the cluster virial radius in addition to the central peak. The background source density is consistent with expectations from {\it COSMOS}. {\bf As in the full band case, the most probable model from our MCMC analysis is overplotted as the black curve ($\beta=-0.52$), and the background source density is statistically consistent with {\it COSMOS}.} }
\end{figure*}

\begin{figure*}
\subfigure[]{
\includegraphics[width=3in, angle=270]{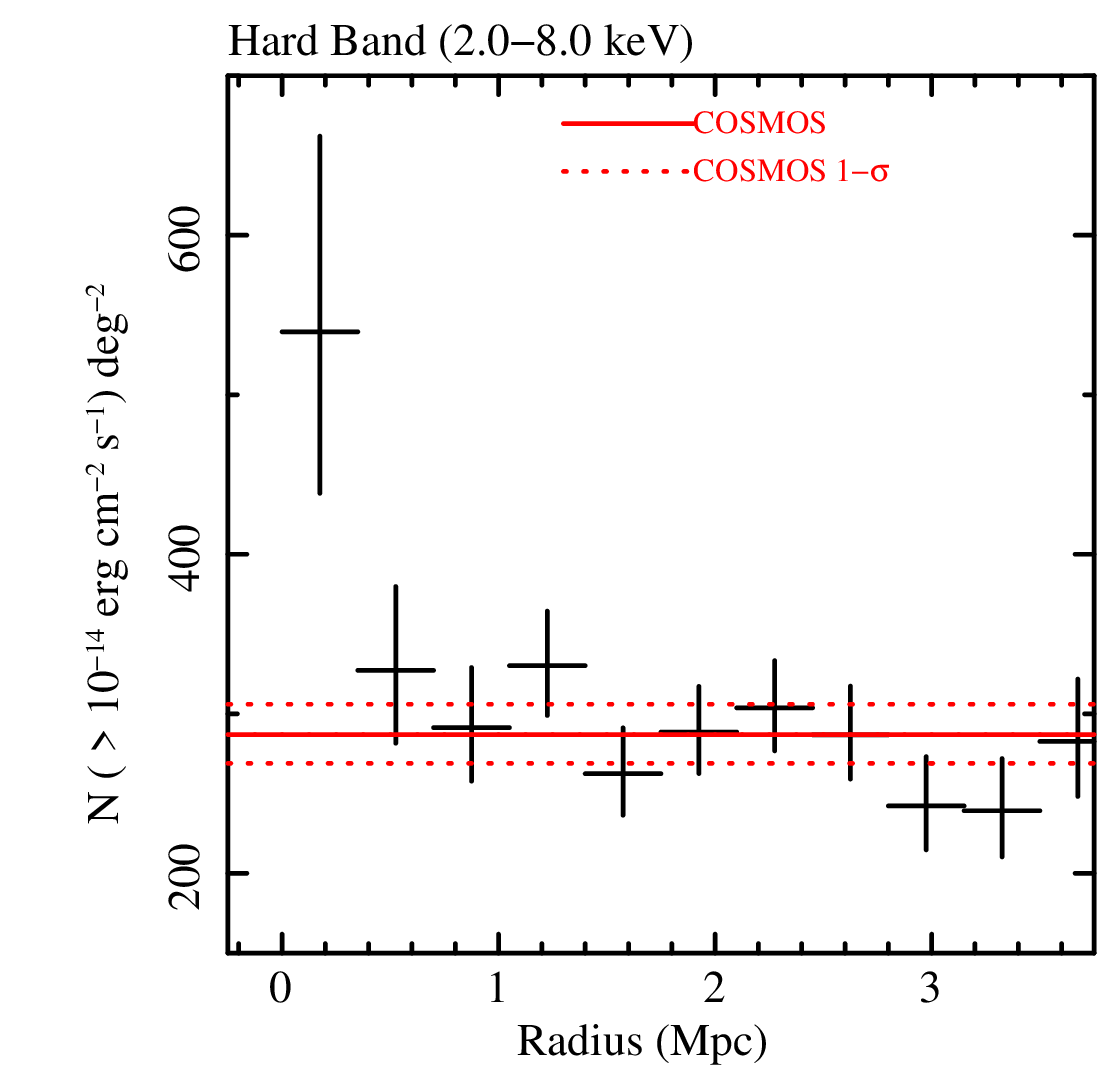}
\label{Mpchardplot}
}
\subfigure[]{
\includegraphics[width=3in, angle=270]{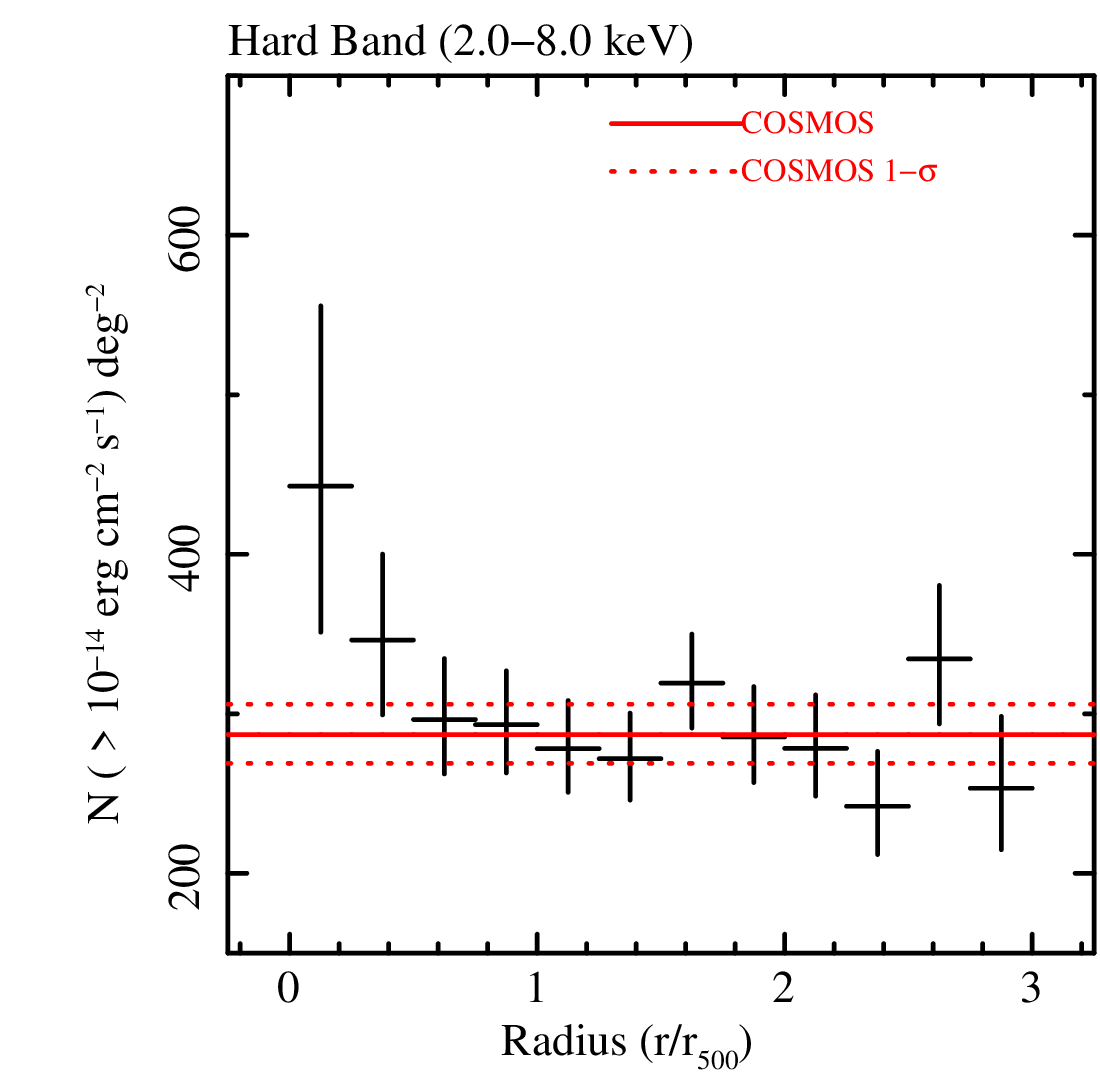}
\label{R500hardplot}
}
\caption{\label{Radhardplots} The projected density of X-ray bright point sources in the soft band ($\mysub{F}{X}(2.0-8.0 \keV) > 10^{-14} \ergpcmsqps$). For both figures, the solid red line denotes the best fit source density of {\it COSMOS} at the corresponding flux limit, while the dashed red lines correspond to its $1-\sigma$ confidence limit. (a): The surface density of X-ray bright hard band sources as a function of radius, in units of \Mpc. A total of 930 sources were included in the calculation of this profile. (b): The surface density of X-ray bright hard band sources as a function of radius, in units of \rfive. Just as in the full and soft bands, using the overdensity radius \rfive \ results in a broader central excess of sources. {\bf The background source density is consistent with expectations from {\it COSMOS} in both profiles. As in the full band case, the most probable model from our MCMC analysis is overplotted as the black curve ($\beta=-0.75$), and the background source density is statistically consistent with {\it COSMOS}.}}
\end{figure*}

Measurements of  the projected source density as a function of cluster-centric radius provide both a powerful investigative tool for understanding the causes of AGN activity and an important systematic cross-check to ensure that detections of excess AGN sources in the cluster fields are not due to fluctuations in the field populations.  The radial distribution of point sources about the cluster centers has been calculated for all point sources with full-band fluxes above $5 \times 10^{-15} \ergpcmsqps$. Similar analyses were performed in the soft band and hard bands, with flux limits of $3 \times 10^{-15} \ergpcmsqps$ and $10^{-14} \ergpcmsqps$, respectively. For all of these flux limits, there is no evidence for survey incompleteness out to $3$\rfive. A total of 2056, 968, and 930 sources satisfy these criteria in the full, soft, and hard bands, respectively. 
The radial distributions are plotted in two ways: firstly, in terms of the metric distance ($r$, in \Mpc, Figure \ref{MPCplot}) from the cluster center;  and secondly (and more usefully), in terms of the overdensity radius ($r$, in \rfive, Figure \ref{R500plot}). The radii (in \Mpc \ or \rfive) of each source were calculated assuming the sources are at the same redshift as the cluster. In both representations, there is evidence of an excess of point sources in the central regions of the clusters. The radial profiles for the soft band and hard band are shown in Figures \ref{R500softplot} and \ref{R500hardplot}, respectively.

At large radii, the measured source number densities converge to a constant source density, consistent with the measured value from \cdfs \ and {\it COSMOS}. Within the virialized cluster region (\rfive) an excess of sources is observed in all three bands. We also observe tentative evidence for a second peak in the profiles between $\sim 1.5-2$ \rfive, especially in the soft band profile. The overall excess of sources above the field population observed in the hard band is smaller than in the soft and full bands, as expected given our hard band flux limit.

Our unique knowledge of the overdensity radii, \rfive, allows for new insights into the influence of the cluster environment on the AGN population. Using Monte Carlo methods, we measure $ 1.1 \pm 0.6$ excess sources per cluster field with respect to {\it COSMOS} within \rfive \ in the full band, $1.1 \pm 0.4$ excess sources per cluster detected within \rfive \ in the soft band, $\sim 0.5 \pm 0.4$ in the hard band.

\begin{table*}
\caption{\label{RadProfResults} Choices of flux limits and expectations from control fields for the radial profiles in all three energy bands. The columns are: (1) The energy band; (2) the flux limit imposed in the radial profile fits, in units of \ergpcmsqps; (3) the density of field sources from the \cdfs \ at that flux limit; (4) the density of field sources from {\it COSMOS} at that flux limit; (5) the number of sources detected within \rfive \ at that flux limit, across all clusters; (6) the survey area within \rfive \ at that flux limit, in units of $\deg^{-2}$; (7) the average excess number of sources per cluster, determined by extrapolating our measurements of the field density from {\it COSMOS} to each radial bin within \rfive; and (8) the best-fit power-law index for the projected source density of cluster member AGN.   }
\centering

\begin{tabular}{ c c c c c c c c }\\
  \hline 

(1) & (2) & (3) & (4) & (5) & (6) & (7) & (8)  \\
Band & Flux Limit ( \ergpcmsqps) & \cdfs ($\deg^{-2}$) & {\it COSMOS} ($\deg^{-2}$)  & $\mysub{n}{500}$ & $\mysub{\Omega}{500}$ ($\deg^{-2}$)& Excess  &  $\beta$  \\ 
\hline\hline
Full & $5 \times 10^{-15}$ & $690 \pm 75$ & $672 \pm 29$  & 565  & 0.77 & $1.1 \pm 0.6$ & $-0.70 \pm 0.23$\\
Soft & $3 \times 10^{-15}$ & $250 \pm 45$ & $255 \pm 18$  &  244 & 0.78 & $1.1 \pm 0.4$ & $-0.52 \pm 0.34$\\
Hard & $1 \times 10^{-14}$ & $220 \pm 45$ & $287 \pm 19$  &  245 & 0.78 & $0.5 \pm 0.4$ & $-0.75 \pm 0.41$\\
&&&&&&&\\
\hline\hline

\end{tabular}
\end{table*}

We have fitted the observed X-ray point source density profiles in all three bands with a Power-Law+Constant model.  

\begin{equation}\label{Powerlaw}
\mysub{N}{X}(r)= \mysub{N}{0} \left(\frac{r}{\mysub{r}{500}}\right)^{\beta} + \mysub{C}{X}
\end{equation}

\noindent We used a Markov Chain Monte Carlo (MCMC) method to determine confidence intervals for the power-law index and accounting for the source count fluctuations in each bin and the covariance between the power-law index and the field density. We assumed a uniform prior for the power-law index $\beta$ between $-5 < \beta <5 $, and sampled over a range of expected field source densities, $\mysub{C}{X}$, using a Gaussian prior based on the {\it COSMOS} value with $\sigma=10\%$. This value for $\sigma$ is sufficiently large to account for both the statistical fluctuations in the {\it COSMOS} source counts ($\sim 5\%$) and the systematic uncertainty due to cosmic variance. The resulting posterior distributions in each energy band are statistically consistent with one another. For the full, soft and hard bands the resulting posterior distributions measure the power-law index as $\beta= -0.70 \pm 0.23$, $\beta= -0.52 \pm 0.34$, and $\beta= -0.75 \pm 0.41$, respectively. The results of fits to all three bands, the expectations from the field surveys, and determinations of the number of excess sources are shown in Table \ref{RadProfResults}.

\subsection{Comparisons with the Cluster Galaxy Distributions}
\begin{figure}
\includegraphics[width=0.92\columnwidth, angle=270]{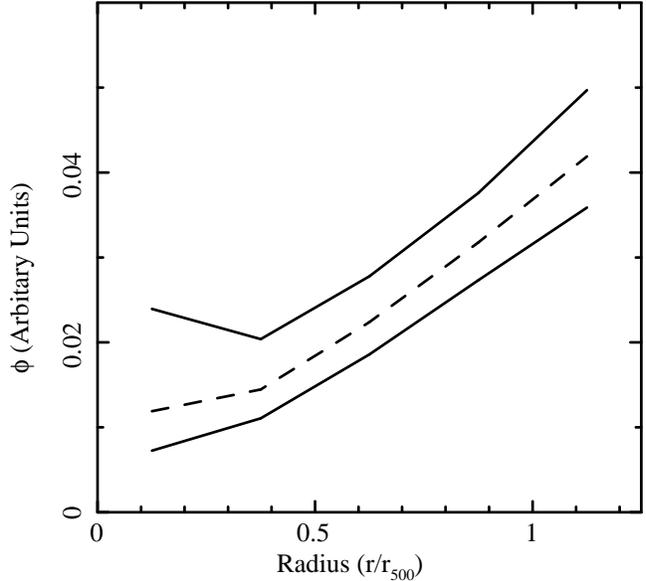}
\caption{\label{Feff} The effective AGN fraction as a function of radius, $\phi(r)$,in the full band ($\mysub{F}{X}(0.5-8.0 \keV) > 5 \times 10^{-15} \ergpcmsqps$). The dashed line corresponds to the median value of $\phi$ in that radial bin, with the solid lines corresponding to the $68\%$ confidence interval. We calculate $\phi(r)$ using an MCMC analysis that determines the most likely power-law model parameters for projected X-ray source density, taking into account all uncertainties in and covariance with the density of field sources. These power-law models were then divided by an assumed model for the projected surface galaxy density \citep{Popesso2007}; a King model with a core radius of $\mysub{r}{c}=0.24 \pm 0.02$ \rfive. The general trends suggest that the AGN fraction at \rfive \ is approximately $\sim 2-4$ times higher than the AGN fraction at $\sim 0.25$\rfive.      }
\end{figure}

Since galaxy clusters are, by definition, regions of high galaxy density, it is likely that any excess of X-ray point sources observed in cluster fields will, at least in part, simply be a result of the overdensity in the underlying galaxy population, rather than influences of the ICM on the evolution of galaxies. Formal counterpart matching between optical and X-ray sources in these clusters, such as those performed in \cite[e.g.][]{Martini2007,Martini2009, Galametz2009}, will be left to future work, but we can nevertheless compare the radial distribution of X-ray sources in our fields with the expected radial distribution of cluster galaxies. Although the AGN fraction cannot be directly determined with such a comparison, we can define an effective AGN fraction, $\phi$ as 
\begin{equation}\label{feffeqn}
\phi(r)=\frac{\mysub{N}{X}(r)-\mysub{C}{X}}{\mysub{N}{O}(r)}
\end{equation}
\noindent where $\mysub{N}{O}(r)$ is a model describing the projected galaxy density, $\mysub{N}{X}(r)$ is the number density of X-ray point sources and $\mysub{C}{X}$ is the background (field) number density. This function should be representative of general trends in the AGN fraction profile, albeit with arbitrary normalization. We perform our analysis within our framework discussed in \S \ref{RadProfSec} to fully account for the covariance between model parameters. We adopt the shape for $\mysub{N}{O}(r)$ from the results of \cite{Popesso2007}, who identified cluster member galaxies for 217 X-ray selected clusters using observations from the Sloan Digital Sky Survey binned by mass, using a magnitude cut of $r < -18.5$. The clusters included in \cite{Popesso2007} span a large range of cluster masses, and like the clusters in this sample were selected based on their X-ray properties. Those authors use a King model parameterization to model the projected galaxy distribution. For the most massive galaxy clusters in their sample ($\geq 5 \times 10^{14} \msolar$, comparable to the masses of clusters in this sample), their model has a core radius of $\mysub{r}{c}=0.24 \pm 0.02$ \rfive. \footnote{We assume that \rtwo / \rfive $=1.5$, which is a good estimate for a large range of concentration parameters for dark matter halos with an NFW density profile \citep{Yang2009}.}  They also fit their projected galaxy data with a projected NFW \citep{Navarro1995,Navarro1997} profile with a concentration parameter $c=\mysub{r}{200}/\mysub{r}{s}=4.2 \pm 0.3$. This result is similar to other studies that use X-ray selected samples of galaxy clusters \citep[e.g.][]{Carlberg1997}. We marginalize over the full range of acceptable King models. Within the framework of this analysis, we divide each realization of the power-law models describing the X-ray sources by these two models for the projected galaxy density, and determine the full posterior distribution for $\phi$ for each radial bin. The median value of the posterior distribution at each radial bin in the full band, along with its corresponding $68\%$ confidence interval is plotted in Figure \ref{Feff}. The resulting profile for $\phi(r)$ is shown to be clearly increasing monotonically with radius. Compared to the value of $\phi$ at $\sim 0.25$\rfive, the data suggest that the AGN fraction is roughly $2-4$ times larger at \rfive. {\bf A similar trend is observed when this same test is performed with the projected NFW models described above. The data presented here do not require any excess sources out beyond \rfive, but a larger sample of galaxy clusters with more additional statistical constraining power will be able to determine whether or not this trend continues out to and beyond the cluster virial radius.   }

\section{Discussion}
We have demonstrated that an excess of X-ray point sources is observed in the directions of 43 massive clusters with \cha, when compared to expectations from deep and medium-deep fields. Although excesses of X-ray point sources in cluster fields have been reported in the literature, our analysis has revealed some details not previously discussed.

The mean excess of sources in these clusters within \rfive \ is $\sim 1$ per cluster field. This is consistent with the results of some previous studies \citep[e.g.][]{Gilmour2009}. These point sources are very likely AGN associated with the clusters, with a typical X-ray luminosity in the full band ($\mysub{L}{X}(0.5-8.0 \keV)$) of $\sim 10^{42-43} \ergps$. 

The projected X-ray AGN source density is highest at the centers of clusters, with excess sources observed out to $\sim $\rfive. The similar excess of sources seen in the soft and full bands given our imposed flux limits ($3 \times 10^{-15} \ergpcmsqps$ in the soft band and $ 5 \times 10^{-15} \ergpcmsqps$ in the full band) suggests that the cluster AGN may have comparable levels of intrinsic obscuration as field AGN. 

 The excess sources associated with the cluster are more broadly distributed than the galaxy distributions in typical X-ray luminous clusters \citep{Popesso2007}. Our initial modeling suggests that, within $\sim \mysub{r}{500}$, the distribution of AGN can be well described by a power-law with an index of $\beta \sim -0.7$. More detailed analysis will require a larger sample and/or improved rejection of non-cluster member X-ray sources. However, we caution that, while the clusters in \cite{Popesso2007} do span a similar mass range to the present sample, they are at lower redshift (mean redshift of $z \sim 0.1$ for \cite{Popesso2007} versus $0.2 < z < 0.7$ for the clusters presented here). Blue galaxies are also generally more broadly distributed than red galaxies. Nonetheless, our initial results suggest that the intrinsic X-ray AGN population in clusters has a broader spatial distribution than both the red and blue cluster member galaxies in \cite{Popesso2007}.\footnote{X-ray AGN in the field are most commonly observed in more massive redder galaxies \citep[e.g.][]{Xue2010}.}  A power-law model of $\beta \sim -0.7$ is broader than nearly all King or projected NFW models with a core/scale radius less than $\sim \mysub{r}{500}$. Future work will explore the properties of the optical counterparts for these X-ray AGN using the \sub \ imaging in hand, providing insights into the host-galaxy properties, and allow us to determine the AGN fraction more robustly. 

We see tentative evidence for a second peak {\bf($\sim 2\sigma$ significance above the {\it COSMOS} density in the soft band)} of sources near the virial radii of the clusters, between $1$ and $2$ \rfive. Further analysis on a larger sample of clusters will be required to confirm the presence of this second peak, given the expected statistical fluctuations for each radial bin. A similar feature was reported by \cite{Ruderman2005}, who utilized soft band ($0.5-2.0 \keV$) measurements and a cluster sample that overlaps with those included here. This feature was not observed in the large cluster sample analysed by \cite{Gilmour2009}, who utilized the full ($0.5-8.0 \keV$) band.

Taking these results at face value, the rising AGN fraction with radius suggests that the evolution of X-ray AGN in galaxy clusters is qualitatively similar to the evolution of of star-forming galaxies. An evolution of the average X-ray AGN fraction in clusters with redshift has been claimed based on comparisons between high and low redshift clusters \citep{Martini2007,Martini2009,Galametz2009}. Our study reveals an effect correlated with cluster radius, which in turn relates to the time since galaxy infall \citep{Gao2004}. Importantly, this trend is qualitatively similar to trends in the fraction of both optically luminous AGN and star-forming galaxies \citep[e.g.][]{Larson1980,Bekki2002,Weinmann2006, Weinmann2009,vandenBosch2008a,vandenBosch2008b, VonderLinden2010}. Although the data presented here do not measure the absolute fraction of cluster galaxies hosting X-ray AGN, they suggest that the ICM may suppress X-ray AGN activity as galaxies are gradually stripped of their fuel reservoirs during infall. Such a result is consistent with a scenario where the X-ray emission of AGN in clusters is linked to the accretion of cold gas.

Our results indicate that the central regions of clusters are relatively inefficient at producing X-ray AGN, in contrast to the presence of radio bright AGN at the centers of many of the clusters in this sample as well as giant elliptical galaxies in general \citep[e.g.][]{Best2004,Best2007,Dunn2008,Dunn2010,Best2012}. Other studies have shown that active cD galaxies are commonly radiatively inefficient \citep[e.g.][]{Dimatteo2001,Dimatteo2003,Best2004,Allen2006,Taylor2006,Best2007}, with radiative emission accounting for $\sim 1\%$ or less of the total AGN energy budget.

Further analyses utilizing a larger sample of X-ray clusters and detailed optical follow-up are underway. Those will provide improved measurements of the radial source density profile, and sufficient data to determine whether the X-ray AGN population in clusters evolves with redshift. Counterpart matching between X-ray sources and optically selected galaxies will be needed to measure the absolute fraction of galaxies hosting X-ray AGN directly and identify which X-ray AGN are cluster members. Measurements of X-ray spectral properties such as the average photon index or X-ray/optical flux ratio offer important clues as to the accretion process for AGN \citep[e.g.][]{Shemmer2005,Shemmer2008,Scott2011}, and probe the impact of the ICM on the accretion process of AGN.

\section{Conclusions}
We have constructed rigorous catalogs of X-ray bright point sources within the fields of 43 massive, X-ray luminous galaxy clusters. We have compared the observed X-ray source population to deep field surveys. Our main results can be summarized as follows:

1)  We measure an excess of point sources detected in the cluster fields when compared with expectations from deep field surveys in the soft ($0.5-2.0 \keV$), hard ($2.0-8.0 \keV$), and full bands ($0.5-8.0 \keV$). On average, this excess amounts to approximately 1 source per cluster field. 

2) The excess AGN are primarily located within $\sim$  \rfive \ of the cluster centers. At large radii, the point source density is in good agreement with expectations from the \cdfs \ and {\it COSMOS} in all three bands.

3) The radial profile for the projected source density can be described by a power-law plus constant model, with $\beta \sim -0.7$, consistently measured across all three energy bands.   

4) A preliminary comparison of our measured AGN profiles with the typical distributions of member galaxies in comparably luminous X-ray selected clusters \citep{Popesso2007} suggests that the fraction of cluster member galaxies hosting X-ray AGN rises with increasing radius. At face value, this suggests that the ICM may suppress X-ray AGN activity as galaxies are gradually stripped of their gas reservoirs during infall, qualitatively similar to the strangulation processes observed for star-forming galaxies in clusters \citep[e.g.][]{VonderLinden2010}. 

Further studies are underway, employing optical counterpart matching and follow-up spectroscopy. This should enable us to investigate the optical properties of the host galaxies, the X-ray spectral properties of the AGN, and provide new insights into the fueling and evolution of X-ray bright AGN in clusters. The full catalog of X-ray sources will be made public at a later date.

\section*{Acknowledgments}
We thank Patrick Broos, Leisa Townsley for technical support, insight, and advice as to how to best utilize the \ae \ software package for this study. We also thank Bret Lehmer for providing the cumulative number counts data for the \cdfs, Nina Bonaventura at the \cha \ X-ray Center's Helpdesk for technical support, and anonymous referee whose comments greatly improved both the results and the discussion. We finally thank Doug Applegate, Patrick Kelly, and Mark Allen for providing preliminary access to the \sub \ data discussed in this paper. Support for this work was provided by the Department of Energy Grant Number DE-AC02-76SF00515 (SE, SWA, AvdL,RGM) and \cha \ X-ray Center grant GO0-11149X. We also acknowledge support from  NASA ADP grant NNX10AC99G (WNB, YQX, BL) and \cha \ X-ray Center grant SP1-12007A (WNB, YQX, BL). We also acknowledge support from the Youth 1000 Plan (QingNianQianRen) program, USTC startup funding (YQX), and National Science Foundation grant AST-0838187 (AM).

\bibliographystyle{mnras}
\input{reference_defs}
\bibliography{PointSources}

\end{document}

%% file: mybeg.tex
\def\spose#1{\hbox to 0pt{#1\hss}}
\def\approxlt{\mathrel{\spose{\lower 3pt\hbox{$\sim$}}
	\raise 2.0pt\hbox{$<$}}}
\def\approxgt{\mathrel{\spose{\lower 3pt\hbox{$\sim$}}
	\raise 2.0pt\hbox{$>$}}}
\def\approxpropto{\mathrel{\spose{\lower 3pt\hbox{$\sim$}}
	\raise 2.0pt\hbox{$\propto$}}}
\mathchardef\twiddle="2218

\def\multleft#1{\hbox to size{\vbox {\halign {\lft{##}\cr #1}}\hfill}\par}
\def\multright#1{\hbox to size{\vbox {\halign {\rt{##}\cr #1}}\hfill}\par}

\def\today{\ifcase\month\or January\or February\or March\or April\or May\or
      June\or July\or August\or September\or October\or November\or December\fi
      \space\number\day, \number\year}
\def\<{\thinspace}

%% file: units.tex
\def\arcsec{{\rm\thinspace arcsec}}
\def\cm{{\rm\thinspace cm}}
\def\erg{{\rm\thinspace erg}}

\def\keV{{\rm\thinspace keV}}

\def\km{{\rm\thinspace km}}
\def\kpc{{\rm\thinspace kpc}}

\def\Mpc{{\rm\thinspace Mpc}}

\def\s{{\rm\thinspace s}}

\def\ergpcmsqps{\hbox{$\erg\cm^{-2}\s^{-1}\,$}}

\def\ergps{\hbox{$\erg\s^{-1}\,$}}

%% file: Table1_new.tex
\begin{table*}
\caption{\label{ChandraSample} Summary of the cluster sample and \cha \ data. A total of 43 clusters of galaxies and 89 independent \cha \ observations were used. The columns are (1) cluster name; (2) redshift; (3) \& (4) right ascension and declination of the cluster X-ray centroid, in degrees (J2000);  (5) the \cha \ observation number (OBS ID\#); (6) observation date; (7) primary detector array used; (8) exposure time in ks, after cleaning; (9) the measured value of the cluster mass $\mysub{M}{500}$, in units of $10^{14} \msolar$; (10) measured value of \rfive \ for the cluster, measured in \Mpc; and (11) the equivalent Galactic hydrogen column density in the direction of the cluster, in units of $10^{20} \rm{atoms} \cm^{-2}$.      }
\centering

\begin{tabular}{ c c c c c c c c c c c}\\

  \hline 

(1) & (2) & (3) & (4) & (5) & (6) & (7) & (8) & (9) & (10) & (11) \\
Cluster Name & {\it z}  & RA &  DEC &  OBS ID \# &  Obs Date  &  Detector  & Exposure (ks) & $\mysub{M}{500} (10^{14} \msolar)$ & \rfive (\Mpc) &  \mysub{N}{H} ($10^{20} \cm^{-2}$)  \\ 
\hline\hline

Abell 209 & 0.206 & $ 22.971 $ & $ -13.613 $  & 3579 & 2003-08-03 & ACIS-I & 9.98 &  12.62 & 1.531 & 1.44\\
... & ... & $ ... $ & $ ... $  & 522 & 2000-09-09 & ACIS-I & 9.96 &  ... & ... & ...\\
Abell 963 & 0.206 & $ 154.264 $ & $ 39.047 $  & 7704 & 2007-02-18 & ACIS-I & 5.06 &  6.82 & 1.247 & 1.25\\
... & ... & $ ... $ & $ ... $  & 903 & 2000-10-11 & ACIS-S & 36.28 &  ... & ... & ...\\
Abell 2261 & 0.224 & $ 260.612 $ & $ 32.132 $  & 5007 & 2004-01-14 & ACIS-I & 24.31 &  14.42 & 1.588 & 3.19\\
... & ... & $ ... $ & $ ... $  & 550 & 1999-12-11 & ACIS-I & 9.06 &  ... & ... & ...\\
Abell 2219 & 0.228 & $ 250.084 $ & $ 46.708 $  & 896 & 2000-03-31 & ACIS-S & 42.29 &  18.87 & 1.737 & 1.76\\
Abell 2390 & 0.233 & $ 328.404 $ & $ 17.695 $  & 4193 & 2003-09-11 & ACIS-S & 95.06 &  15.15 & 1.613 & 6.21\\
... & ... & $ ... $ & $ ... $  & 500 & 2000-10-08 & ACIS-S & 9.82 &  ... & ... & ...\\
... & ... & $ ... $ & $ ... $  & 501 & 1999-11-05 & ACIS-S & 9.04 &  ... & ... & ...\\
RXJ2129.6+0005 & 0.235 & $ 322.415 $ & $ 0.088 $  & 552 & 2000-10-21 & ACIS-I & 9.96 &  7.71 & 1.285 & 3.63\\
... & ... & $ ... $ & $ ... $  & 9370 & 2009-04-03 & ACIS-I & 29.64 &  ... & ... & ...\\
Abell 521 & 0.248 & $ 73.530 $ & $ -10.223 $  & 430 & 2000-10-13 & ACIS-S & 39.11 &  11.43 & 1.459 & 4.78\\
... & ... & $ ... $ & $ ... $  & 901 & 1999-12-23 & ACIS-I & 38.63 &  ... & ... & ...\\
Abell 1835 & 0.253 & $ 210.258 $ & $ 2.877 $  & 6880 & 2006-08-25 & ACIS-I & 117.91 &  12.29 & 1.493 & 2.04\\
... & ... & $ ... $ & $ ... $  & 6881 & 2005-12-07 & ACIS-I & 36.28 &  ... & ... & ...\\
... & ... & $ ... $ & $ ... $  & 7370 & 2006-07-24 & ACIS-I & 39.50 &  ... & ... & ...\\
Abell 68 & 0.255 & $ 9.274 $ & $ 9.160 $  & 3250 & 2002-09-07 & ACIS-I & 9.98 &  7.61 & 1.270 & 4.96\\
Abell 697 & 0.282 & $ 130.740 $ & $ 36.365 $  & 4217 & 2002-12-15 & ACIS-I & 19.51 &  17.10 & 1.647 & 2.93\\
Abell 2537 & 0.297 & $ 347.091 $ & $ -2.191 $  & 4962 & 2004-09-09 & ACIS-S & 36.19 &  7.15 & 1.225 & 4.62\\
... & ... & $ ... $ & $ ... $  & 9372 & 2008-08-11 & ACIS-I & 38.50 &  ... & ... & ...\\
MS2137.3-2353 & 0.313 & $ 325.063 $ & $ -23.661 $  & 4974 & 2003-11-13 & ACIS-S & 57.38 &  4.74 & 1.063 & 3.76\\
... & ... & $ ... $ & $ ... $  & 5250 & 2003-11-18 & ACIS-S & 40.54 &  ... & ... & ...\\
... & ... & $ ... $ & $ ... $  & 928 & 1999-11-18 & ACIS-S & 43.60 &  ... & ... & ...\\
MACSJ1931.8-2634 & 0.352 & $ 292.956 $ & $ -26.576 $  & 3282 & 2002-10-20 & ACIS-I & 13.59 &  9.94 & 1.339 & 8.31\\
... & ... & $ ... $ & $ ... $  & 9382 & 2008-08-21 & ACIS-I & 98.92 &  ... & ... & ...\\
MACSJ1115.8+0129 & 0.355 & $ 168.966 $ & $ 1.498 $  & 3275 & 2003-01-23 & ACIS-I & 15.90 &  8.64 & 1.278 & 4.34\\
... & ... & $ ... $ & $ ... $  & 9375 & 2008-02-03 & ACIS-I & 39.62 &  ... & ... & ...\\
RXJ1532.9+3021 & 0.363 & $ 233.224 $ & $ 30.349 $  & 1649 & 2001-08-26 & ACIS-S & 9.36 &  9.48 & 1.312 & 2.30\\
... & ... & $ ... $ & $ ... $  & 1665 & 2001-09-06 & ACIS-I & 9.97 &  ... & ... & ...\\
ZWCL1953 & 0.378 & $ 132.529 $ & $ 36.072 $  & 1659 & 2000-10-22 & ACIS-I & 24.86 &  10.16 & 1.336 & 2.96\\
... & ... & $ ... $ & $ ... $  & 7716 & 2006-12-20 & ACIS-I & 6.98 &  ... & ... & ...\\
MACSJ0949.8+1708 & 0.384 & $ 147.465 $ & $ 17.118 $  & 3274 & 2002-11-06 & ACIS-I & 14.31 &  11.35 & 1.380 & 3.08\\
MACSJ1720.2+3536 & 0.387 & $ 260.069 $ & $ 35.606 $  & 3280 & 2002-11-03 & ACIS-I & 20.84 &  6.31 & 1.135 & 3.46\\
... & ... & $ ... $ & $ ... $  & 6107 & 2005-11-22 & ACIS-I & 33.88 &  ... & ... & ...\\
MACSJ1731.6+2252 & 0.389 & $ 262.913 $ & $ 22.863 $  & 3281 & 2002-11-03 & ACIS-I & 20.50 &  12.82 & 1.427 & 4.99\\
MACSJ2211.7-0349 & 0.396 & $ 332.941 $ & $ -3.828 $  & 3284 & 2002-10-08 & ACIS-I & 17.73 &  18.06 & 1.608 & 5.53\\
MACSJ0429.6-0253 & 0.399 & $ 67.400 $ & $ -2.884 $  & 3271 & 2002-02-07 & ACIS-I & 23.16 &  5.76 & 1.097 & 4.34\\
MACSJ2228.5+2036 & 0.411 & $ 337.136 $ & $ 20.620 $  & 3285 & 2003-01-22 & ACIS-I & 19.85 &  14.67 & 1.491 & 4.26\\
MACSJ0451.9+0006 & 0.429 & $ 72.977 $ & $ 0.105 $  & 5815 & 2005-01-08 & ACIS-I & 10.21 &  6.33 & 1.118 & 6.85\\
MACSJ1206.2-0847 & 0.439 & $ 181.551 $ & $ -8.801 $  & 3277 & 2002-12-15 & ACIS-I & 23.45 &  19.16 & 1.612 & 4.35\\
MACSJ0417.5-1154 & 0.443 & $ 64.393 $ & $ -11.907 $  & 11759 & 2009-10-28 & ACIS-I & 51.35 &  22.06 & 1.689 & 3.31\\
... & ... & $ ... $ & $ ... $  & 12010 & 2009-10-29 & ACIS-I & 25.78 &  ... & ... & ...\\
... & ... & $ ... $ & $ ... $  & 3270 & 2002-03-10 & ACIS-I & 12.01 &  ... & ... & ...\\
MACSJ2243.3-0935 & 0.447 & $ 340.839 $ & $ -9.595 $  & 3260 & 2002-12-23 & ACIS-I & 20.50 &  17.35 & 1.555 & 4.02\\

&&&&&&&& \\
\hline

\end{tabular}
\end{table*}

\addtocounter{table}{-1}

\begin{table*}
\caption{Continued  }
\centering

\begin{tabular}{ c c c c c c c c c c c}\\

  \hline
(1) & (2) & (3) & (4) & (5) & (6) & (7) & (8) & (9) & (10) & (11)\\
 Cluster Name & {\it z}  &  RA &  DEC &  OBS ID \# &  Obs Date  &  Detector  &  Exposure (ks) &  $\mysub{M}{500} (10^{14} \msolar)$ & \rfive (\Mpc) &  \mysub{N}{H} ($10^{20} \cm^{-2}$)  \\ 
\hline\hline

MACSJ0329.6-0211 & 0.450 & $ 52.422 $ & $ -2.195 $  & 3257 & 2001-11-25 & ACIS-I & 9.86 &  7.89 & 1.194 & 4.64\\
... & ... & $ ... $ & $ ... $  & 3582 & 2002-12-24 & ACIS-I & 19.84 &  ... & ... & ...\\
... & ... & $ ... $ & $ ... $  & 6108 & 2004-12-06 & ACIS-I & 39.64 &  ... & ... & ...\\
... & ... & $ ... $ & $ ... $  & 7719 & 2006-12-03 & ACIS-I & 7.08 &  ... & ... & ...\\
RXJ1347.5-1145 & 0.451 & $ 206.878 $ & $ -11.752 $  & 3592 & 2003-09-03 & ACIS-I & 57.71 &  21.71 & 1.674 & 4.60\\
... & ... & $ ... $ & $ ... $  & 506 & 2000-03-05 & ACIS-S & 8.93 &  ... & ... & ...\\
... & ... & $ ... $ & $ ... $  & 507 & 2000-04-29 & ACIS-S & 9.99 &  ... & ... & ...\\
MACSJ1621.3+3810 & 0.463 & $ 245.353 $ & $ 38.169 $  & 10785 & 2008-10-18 & ACIS-I & 29.75 &  5.89 & 1.078 & 1.13\\
... & ... & $ ... $ & $ ... $  & 3254 & 2002-10-18 & ACIS-I & 9.84 &  ... & ... & ...\\
... & ... & $ ... $ & $ ... $  & 6109 & 2004-12-11 & ACIS-I & 37.54 &  ... & ... & ...\\
... & ... & $ ... $ & $ ... $  & 6172 & 2004-12-25 & ACIS-I & 29.75 &  ... & ... & ...\\
... & ... & $ ... $ & $ ... $  & 9379 & 2008-10-17 & ACIS-I & 29.91 &  ... & ... & ...\\
MACSJ1108.8+0906 & 0.466 & $ 167.229 $ & $ 9.100 $  & 3252 & 2002-11-17 & ACIS-I & 9.94 &  7.73 & 1.179 & 2.22\\
... & ... & $ ... $ & $ ... $  & 5009 & 2004-02-20 & ACIS-I & 24.46 &  ... & ... & ...\\
MACSJ1427.2+4407 & 0.487 & $ 216.816 $ & $ 44.125 $  & 6112 & 2005-02-12 & ACIS-I & 9.38 &  6.35 & 1.095 & 1.19\\
... & ... & $ ... $ & $ ... $  & 9380 & 2008-01-14 & ACIS-I & 25.81 &  ... & ... & ...\\
... & ... & $ ... $ & $ ... $  & 9808 & 2008-01-15 & ACIS-I & 14.93 &  ... & ... & ...\\
MACSJ2214.9-1359 & 0.502 & $ 333.738 $ & $ -14.003 $  & 3259 & 2002-12-22 & ACIS-I & 19.47 &  13.16 & 1.387 & 2.88\\
... & ... & $ ... $ & $ ... $  & 5011 & 2003-11-17 & ACIS-I & 18.52 &  ... & ... & ...\\
MACSJ0911.2+1746 & 0.505 & $ 137.795 $ & $ 17.775 $  & 3587 & 2003-02-23 & ACIS-I & 17.87 &  8.96 & 1.220 & 3.28\\
... & ... & $ ... $ & $ ... $  & 5012 & 2004-03-08 & ACIS-I & 23.79 &  ... & ... & ...\\
MACSJ0257.1-2325 & 0.505 & $ 44.287 $ & $ -23.434 $  & 1654 & 2000-10-03 & ACIS-I & 19.84 &  8.51 & 1.198 & 2.08\\
... & ... & $ ... $ & $ ... $  & 3581 & 2003-08-23 & ACIS-I & 18.47 &  ... & ... & ...\\
MACSJ0454.1-0300 & 0.538 & $ 73.547 $ & $ -3.014 $  & 529 & 2000-01-14 & ACIS-I & 13.90 &  11.46 & 1.308 & 3.92\\
... & ... & $ ... $ & $ ... $  & 902 & 2000-10-08 & ACIS-S & 44.19 &  ... & ... & ...\\
MACSJ1423.8+2404 & 0.543 & $ 215.949 $ & $ 24.078 $  & 1657 & 2001-06-01 & ACIS-I & 18.52 &  6.65 & 1.088 & 2.20\\
... & ... & $ ... $ & $ ... $  & 4195 & 2003-08-18 & ACIS-S & 115.57 &  ... & ... & ...\\
MACSJ1149.5+2223 & 0.544 & $ 177.397 $ & $ 22.401 $  & 1656 & 2001-06-01 & ACIS-I & 18.52 &  18.66 & 1.533 & 1.92\\
... & ... & $ ... $ & $ ... $  & 3589 & 2003-02-07 & ACIS-I & 20.04 &  ... & ... & ...\\
MACSJ0717.5+3745 & 0.546 & $ 109.383 $ & $ 37.755 $  & 1655 & 2001-01-29 & ACIS-I & 19.87 &  24.87 & 1.688 & 6.64\\
... & ... & $ ... $ & $ ... $  & 4200 & 2003-01-08 & ACIS-I & 59.16 &  ... & ... & ...\\
MS0015.9+1609 & 0.547 & $ 4.639 $ & $ 16.436 $  & 520 & 2000-08-18 & ACIS-I & 67.41 &  16.47 & 1.469 & 3.99\\
MACSJ0025.4-1222 & 0.585 & $ 6.374 $ & $ -12.379 $  & 10413 & 2008-10-16 & ACIS-I & 75.63 &  7.59 & 1.119 & 2.50\\
... & ... & $ ... $ & $ ... $  & 10786 & 2008-10-18 & ACIS-I & 14.12 &  ... & ... & ...\\
... & ... & $ ... $ & $ ... $  & 10797 & 2008-10-21 & ACIS-I & 23.85 &  ... & ... & ...\\
... & ... & $ ... $ & $ ... $  & 3251 & 2002-11-11 & ACIS-I & 19.32 &  ... & ... & ...\\
... & ... & $ ... $ & $ ... $  & 5010 & 2004-08-09 & ACIS-I & 24.82 &  ... & ... & ...\\
MACSJ2129.4-0741 & 0.588 & $ 322.357 $ & $ -7.691 $  & 3199 & 2002-12-23 & ACIS-I & 19.85 &  10.65 & 1.250 & 4.33\\
... & ... & $ ... $ & $ ... $  & 3595 & 2003-10-18 & ACIS-I & 19.87 &  ... & ... & ...\\
MACSJ0647.7+7015 & 0.592 & $ 101.957 $ & $ 70.248 $  & 3196 & 2002-10-31 & ACIS-I & 19.27 &  10.88 & 1.257 & 5.40\\
... & ... & $ ... $ & $ ... $  & 3584 & 2003-10-07 & ACIS-I & 19.99 &  ... & ... & ...\\
MACSJ0744.8+3927 & 0.698 & $ 116.217 $ & $ 39.457 $  & 3197 & 2001-11-12 & ACIS-I & 20.23 &  12.53 & 1.264 & 5.66\\
... & ... & $ ... $ & $ ... $  & 3585 & 2003-01-04 & ACIS-I & 19.85 &  ... & ... & ...\\
... & ... & $ ... $ & $ ... $  & 6111 & 2004-12-03 & ACIS-I & 49.50 &  ... & ... & ...\\

&&&&&&&& \\
\hline

\end{tabular}
\end{table*}

%% file: Table2_new.tex
\begin{table}
\caption{\label{ChandraProperties}  Summary of AGN source numbers and flux limits for each cluster field. The columns are (1) cluster name; (2) the number of sources in the initial candidate point source catalog for each cluster, produced by \wav; (3) the number of sources included in the final catalog followed by the number that satisfy \thresh \ in the full band, the soft band, and the hard band, respectively; (4) the flux limit for each cluster observation in the full, soft, and hard bands, defined as the minimum flux to which 50\% of the survey area is sensitive, in units of $10^{-15} \ergpcmsqps$. Clusters denoted with a dagger ($\dagger$) utilized a mixture of ACIS-I and ACIS-S observations.  }
\centering
\begin{tabular}{ c c c c}\\
  \hline Cluster Name  & \mysub{N}{wav} & \mysub{N}{cat}/\mysub{N}{full}/\mysub{N}{soft}/\mysub{N}{hard} &  Flux Limit  \\ 
\hline\hline

Abell 209 & 155 & 56/52/37/34 & 5.01/2.00/6.31\\
Abell 963$^{\dagger}$ & 122 & 79/77/63/47 & 8.91/2.51/14.13\\
Abell 2261 & 117 & 75/72/62/50 & 3.98/1.58/5.62\\
Abell 2219 & 59 & 40/35/32/18 & 3.16/1.26/4.47\\
Abell 2390 & 146 & 93/93/74/56 & 5.62/1.58/10.00\\
RXJ2129.6+0005 & 282 & 109/104/85/59 & 5.01/1.78/7.08\\
Abell 521$^{\dagger}$ & 185 & 141/137/113/84 & 3.16/1.00/4.47\\
Abell 1835 & 220 & 177/174/139/116 & 2.51/0.79/3.16\\
Abell 68 & 81 & 50/49/38/31 & 7.08/2.82/10.00\\
Abell 697 & 156 & 69/67/52/43 & 5.01/1.78/6.31\\
Abell 2537$^{\dagger}$ & 139 & 104/102/81/64 & 3.55/1.12/5.01\\
MS2137.3-2353 & 105 & 68/66/54/44 & 1.78/0.56/2.82\\
MACSJ1931.8-2634 & 162 & 137/135/105/93 & 1.78/0.63/2.51\\
MACSJ1115.8+0129 & 109 & 94/93/77/63 & 2.51/0.89/3.55\\
RXJ1532.9+3021$^{\dagger}$ & 244 & 46/43/38/26 & 6.31/2.00/7.08\\
ZWCL1953 & 145 & 106/103/85/52 & 3.98/1.41/5.62\\
MACSJ0949.8+1708 & 96 & 49/48/39/34 & 5.62/2.24/7.08\\
MACSJ1720.2+3536 & 150 & 125/121/102/83 & 2.82/0.89/3.55\\
MACSJ1731.6+2252 & 136 & 83/80/65/49 & 4.47/1.58/5.62\\
MACSJ2211.7-0349 & 144 & 80/77/55/46 & 5.01/1.78/7.08\\
MACSJ0429.6-0253 & 147 & 91/89/70/59 & 3.98/1.41/5.62\\
MACSJ2228.5+2036 & 128 & 78/76/61/51 & 4.47/1.58/6.31\\
MACSJ0451.9+0006 & 84 & 52/51/40/31 & 7.08/2.82/10.00\\
MACSJ1206.2-0847 & 176 & 86/84/59/53 & 3.98/1.41/5.01\\
MACSJ0417.5-1154 & 163 & 131/127/102/87 & 4.47/1.00/4.47\\
MACSJ2243.3-0935 & 126 & 62/59/49/38 & 4.47/1.78/5.62\\
MACSJ0329.6-0211 & 132 & 102/96/77/63 & 2.51/0.89/3.55\\
RXJ1347.5-1145$^{\dagger}$ & 366 & 86/81/60/57 & 2.82/1.00/5.01\\
MACSJ1621.3+3810 & 186 & 159/154/134/112 & 2.00/0.71/2.82\\
MACSJ1108.8+0906 & 173 & 80/79/60/47 & 3.55/1.26/5.01\\
MACSJ1427.2+4407 & 186 & 95/95/74/62 & 2.82/1.00/3.98\\
MACSJ2214.9-1359 & 154 & 104/104/83/60 & 3.16/1.12/4.47\\
MACSJ0911.2+1746 & 103 & 88/85/75/51 & 2.82/1.00/3.98\\
MACSJ0257.1-2325 & 164 & 106/102/85/58 & 3.55/1.26/5.01\\
MACSJ0454.1-0300 & 169 & 85/80/70/48 & 6.31/2.24/7.94\\
MACSJ1423.8+2404$^{\dagger}$ & 185 & 127/123/105/84 & 4.47/1.41/7.08\\
MACSJ1149.5+2223$^{\dagger}$ & 129 & 107/105/85/66 & 6.31/2.24/7.94\\
MACSJ0717.5+3745 & 200 & 156/146/133/95 & 2.82/1.00/3.98\\
MS0015.9+1609 & 140 & 116/114/94/81 & 1.78/0.63/2.51\\
MACSJ0025.4-1222 & 203 & 182/176/150/113 & 2.82/0.89/4.47\\
MACSJ2129.4-0741 & 129 & 106/104/88/67 & 3.16/1.12/4.47\\
MACSJ0647.7+7015 & 106 & 91/89/68/50 & 3.16/1.12/4.47\\
MACSJ0744.8+3927 & 159 & 139/137/111/98 & 2.24/0.71/3.16\\

&&& \\
\hline

\end{tabular}
\end{table}

%% file: reference_defs.tex
\def \aap {A\&A} 
\def \statisci {Statis. Sci.}
\def \physrep {Phys. Rep.}
\def \pre {Phys.\ Rev.\ E}
\def \sjos {Scand. J. Statis.} 
\def \jrssb {J. Roy. Statist. Soc. B} 


%

\def \araa {ARA\&A}
\def \aj {AJ}
 \def \aas {A\&AS}
\def \apj {ApJ}
\def \apjl {ApJL}
\def \apjs {ApJS}
\def \mnras {MNRAS}
\def \nat {Nat}
 \def \pasp {PASP}
\def \gca {Geochim.\ Cosmochim.\ Acta}
\def \prd {Phys.\ Rev.\ D}
\def \prl {Phys.\ Rev.\ Lett.}